\documentclass[aps,prd,superscriptaddress,nofootinbib,eqsecnum,twocolumn]{revtex4-2}

\pdfoutput=1

\usepackage{amsfonts}
\usepackage{amsmath}
\usepackage{amssymb}
\usepackage{graphicx,color}
\usepackage{float}
\usepackage{hyperref}
\usepackage{subfigure}
\usepackage{dcolumn}
\usepackage{soul}
\usepackage{ulem}
\usepackage{graphicx}
\usepackage{color}

\begin{document}

\title{Scale dependence improvement of the quartic scalar field thermal effective potential in the optimized perturbation theory}
 
\author{Lucas G. C\^amara} \email{lucasgondim@if.ufrj.br}
\affiliation{Departamento de Física Teórica, Universidade do Estado do
  Rio de Janeiro, Rio de Janeiro, RJ, Brazil}

\author{Marcus Benghi Pinto} \email{ marcus.benghi@ufsc.br}
\affiliation{Departamento de Física, Universidade Federal de Santa Catarina, Florianópolis, SC, Brazil}
\affiliation{Laboratoire Charles Coulomb (L2C), UMR 5221 CNRS-Universit\'{e} Montpellier, 34095 Montpellier, France}

\author{Rudnei O. Ramos} \email{rudnei@uerj.br}
\affiliation{Departamento de Física Teórica, Universidade do Estado do
  Rio de Janeiro, Rio de Janeiro, RJ, Brazil}

\begin{abstract}

Perturbation theory,   as well as most thermal field resummation methods widely used to study finite-temperature quantum field theories, presents a non-negligible renormalization scale dependence. To address this limitation, we propose an alternative method that combines the renormalization group improvement prescription for the thermal effective potential with the optimized perturbation theory variational resummation technique. 
Here, we apply this new framework, termed variational renormalization group, to evaluate the effective potential of the scalar $\lambda \phi^4$ theory at finite temperatures, which represents a benchmark model for phase transition studies. We show that the proposed approach significantly improves scale stability, compared to the use of optimized perturbation theory alone, across key thermodynamic quantities, including the effective potential, critical temperature, and pressure. 
These results establish the variational renormalization group as a robust alternative tool for precision studies of thermal phase transitions, with direct implications for cosmological applications (e.g., early-Universe thermodynamics) and condensed matter systems. 

\end{abstract}

\maketitle

\section{Introduction}

In quantum field theories at finite temperatures, conventional perturbation theory often fails due to the presence of infrared (IR) divergences and the breakdown of the perturbative expansion at high temperatures~\cite{Kapusta:2006pm}. Thermal effects introduce a new energy scale, represented by the temperature \( T \), which amplifies long-wavelength (soft) fluctuations. For example, in scalar \(\lambda \phi^4\) theory, loop corrections to the effective potential acquire terms proportional to \(\lambda T^2\), even for small coupling \(\lambda\). When \( \lambda T^2 \) becomes large (e.g., near a phase transition), higher-order terms in the perturbative series (e.g., \(\lambda^n T^{2n}\)) grow uncontrollably, destroying convergence. This is exacerbated by the appearance of IR divergences in contributions with repeated soft-momentum exchanges (e.g., ``ring diagrams"), which diverge as \( \int d^3k/k^2 \) at finite \( T \), rendering naive perturbation theory ill defined.  

To address this issue, resummation techniques, e.g., hard thermal loop (HTL) resummation and similar techniques~\cite{Andersen:2000yj,Andersen:2004fp,Su:2012iy}, reorganize the perturbative series by selectively summing infinite classes of dominant diagrams, such as those encoding Debye screening of IR singularities. {}For example, HTL resummation incorporates thermal masses \( m_{\text{th}}^2 \sim \lambda T^2 \) into propagators, taming IR divergences, and restoring a controlled expansion.  Hence, resummation and nonperturbative tools are indispensable for modeling finite-temperature systems, ranging from early-Universe cosmology to quark-gluon plasma physics.

Although resummation methods in thermal field theories can successfully mitigate infrared divergences and restore perturbative control in finite-temperature quantum field theory, they leave unresolved the significant renormalization-scale dependence of thermodynamic quantities like the effective potential, pressure, and critical temperature. This residual scale sensitivity arises because the resummation process reorganizes the perturbative series without fully enforcing renormalization group (RG) invariance, a fundamental property of physical observables~\cite{Lofgren:2023sep}. For example, within scalar \(\lambda \phi^4\) theory, the resummed effective potential \(V_{\text{\rm eff}}(T, \mu)\) depends on the arbitrary renormalization scale \(\mu\), with variations of \(\mu\) by a factor of 2 often inducing more than \(\sim 20-30\%\) changes in calculated quantities such as the critical temperature \(T_c\). 
In general, to circumvent the  scale dependence problem in resummation evaluations, it is common to adopt a range of values for the renormalization scale \(\mu\), presenting results within the range \(\mu \in \left[\pi T, 4\pi T \right]\). However, even when this range of energy scale values is adopted, one still obtains a strong scale dependence. This behavior seems somewhat counterintuitive, since a fourfold variation in the energy scale should not lead to drastic effects in physical quantities such as the pressure, for example. Moreover, these results worsen as the perturbative order increases \cite{Andersen:2009tc,Andersen:2011sf,Mogliacci:2013mca,Haque:2013sja,Bandyopadhyay:2015gva}.
Such ambiguities undermine precision in applications such as those related to early-Universe phase transition~\cite{Croon:2020cgk,Athron:2023xlk,Hashino:2025nku}, where, for example, predictions for bubble nucleation rates or gravitational wave spectra depend sensitively on the critical temperature \(T_c\).  

In this situation, one may recur to Renormalization Group Improvement (RGI) techniques~\cite{Gould:2021dzl,Gould:2021oba,Funakubo:2023eic}  and modify the resummed expressions by imposing RG invariance through the incorporation of logarithmic terms (\(\sim \ln(\mu/T)\)) dictated by the \(\beta\) functions and anomalous dimensions of the theory.  However, even RGI cannot fully eliminate scale sensitivity in strongly coupled regimes (\( \lambda T^2 \gtrsim \mathcal{O}(1) \)), where nonperturbative effects dominate. Thus, a hybrid approach (e.g. combining resummation methods and RGI) is essential to produce an accurate description of the thermodynamics associated with physical systems ranging from electroweak symmetry breaking to quark-gluon plasma dynamics. 

In the present work, we propose a hybrid approach that combines the optimized perturbation theory (OPT) variational resummation method with the renormalization group technique RGI. This framework~\cite{Okopinska:1987hp,Duncan:1988hw} (see also Ref.~\cite{Yukalov:2019nhu} for a recent review)  was conceived to improve the convergence of perturbative expansions, which often suffer from divergence or poor behavior at strong couplings. In standard perturbation theory, physical quantities are expanded as a power series in the coupling constant, but such series are typically asymptotic and may not yield accurate results beyond leading orders. The method addresses this issue by modifying the original Lagrangian: It introduces an artificial mass parameter ($\eta$) in such a way that the modified theory interpolates between a solvable (noninteracting) theory and the full interacting theory. The key step is to perform a perturbative expansion around this modified theory and then optimize the result by fixing $\eta$ so that the physical quantity of interest is least sensitive to variations in it; this is known as the principle of minimal sensitivity (PMS). This procedure allows for more reliable approximations even in regimes where standard perturbation theory fails, making it valuable for studying nonperturbative aspects of quantum field systems.
The OPT has already been applied to a plethora of
problems, including low energy
systems of interest in condensed matter ~\cite{deSouzaCruz:2000fy,Caldas:2008zz,Caldas:2009zz,Gomes:2023vvu},
to the study of chiral phase transition in QCD effective
models~\cite{Kneur:2010yv,Kneur:2012qp,Restrepo:2014fna,Duarte:2017zdz}
and also to different problems in thermal quantum field theory~\cite{Klimenko:1992av,Pinto:1999py,Pinto:1999pg,Kneur:2007vj,Kneur:2007vm,Farias:2008fs,Farias:2021ult,Silva:2023jrk,Martins:2024dag,Tavares:2024edx}. The method shows a fast
convergence~\cite{Kneur:2002kq,Rosa:2016czs}, and already at the first nontrivial order it is able to produce
results improving over other nonperturbative methods, e.g., the
large-$N$ expansion and Gaussian approximations, becoming equivalent
to the daisy and superdaisy nonperturbative schemes~\cite{Duarte:2011ph}.

In the context of OPT, a new scheme called renormalization group optimized perturbation theory (RGOPT) was developed to address the scale dependence problem~\cite{Kneur:2010ss,Kneur:2011zz,Kneur:2013coa,Kneur:2015dda,Kneur:2015moa,Kneur:2015uha,Fernandez:2021sgr,Restrepo:2025qgp}.
This method modifies the original OPT implementation and introduces a new (subtraction) term into the free energy, ensuring invariance under the renormalization group. The traditional OPT mass term, in a scalar field theory, $(1-\delta)\eta^2$, is written as $(1-\delta)^a\eta^2$ where $a$ turns out to be a function of the RG $\beta$ and $\gamma_m$ coefficients $b_0$ and $\gamma_0$. These adaptations have allowed this approximation to achieve great success in investigating the symmetric phase of several models, including QCD, as mentioned earlier. However, so far the method has not yet been used, for example, in the description of phase transitions. The approach that we propose in the present paper differs fundamentally from the RGOPT by preserving the
original prescription of the OPT method and improving over the scale dependence displayed by the thermodynamic quantities by
applying directly the OPT resummation over the RGI improved effective potential. We termed this alternative approach the variational renormalization group (VRG).
We will see that by preserving the original OPT method, augmented with RG invariance properties, we can study the
system's thermodynamic properties in both the symmetric and broken phases in a systematic way.

This work is organized as follows. In Sec.~\ref{sec2}, we briefly review the application of the OPT 
method to our fiducial $\lambda \phi^4$ scalar field theory.
The effective potential at finite temperature up
to second order in the OPT is explicitly written down.
In Sec.~\ref{sec3}, we review how the RGI method allows an improvement of the effective potential with 
respect to scale variations. 
We  also show in that section how the RGI method can be modified to carry out an improvement in the 
context of the OPT finite-temperature effective potential, ensuring that the improvement remains 
consistent with the renormalization group.
In Sec.~\ref{sec4}, we present our main numerical results for the thermodynamics of the quartic scalar field when using the VRG.
{}Finally, in Sec.~\ref{conclusions}, we draw our main conclusions and perspectives for future work.
{}Five appendixes are also included where the relevant definitions, expressions and technical details used in our analysis are given.

\section{The effective potential of
the $\lambda \phi^4$ model in the OPT at next-to-leading order}
\label{sec2}

The OPT implementation consists of an interpolation procedure followed by the application of variational condition that generates optimal (nonperturbative) results.
In our case, we start from the standard Lagrangian density describing the $\lambda \phi^4$ theory:
\begin{eqnarray}\label{Lagr}
    \mathcal{L} &=& \frac{1}{2}(\partial_\mu \phi)(\partial^\mu \phi) - \frac{m^2}{2} \phi^2  - \frac{\lambda}{4!} \phi^4.
\end{eqnarray}
The OPT prescription starts by implementing in the Lagrangian density the following replacements~\cite {Pinto:1999py,Pinto:1999pg,Kneur:2007vj,Kneur:2007vm}
\begin{eqnarray}
        m^2 &\to & m^2 + (1-\delta) \eta^2, \label{Mmodopt} \\
        \lambda &\to & \delta \lambda \label{Lmodopt} \,,
\end{eqnarray}
obtaining the OPT deformed Lagrangian density
\begin{eqnarray}\label{LrLct}
    \mathcal{L}^{\delta} &=& \frac{1}{2}(\partial_\mu \phi)(\partial^\mu \phi) - \frac{\Omega^2}{2} \phi^2 + \frac{\delta \eta^2}{2} \phi^2 - \frac{\delta \lambda}{4!} \phi^4,
\nonumber \\
\end{eqnarray}
where $\delta$ represents an artificial bookkeeping parameter (formally considered to be small), while $\Omega^2= m^2 + \eta^2$ contains the important 
 arbitrary variational mass parameter, $\eta$. The quantities of interest are evaluated
up to some order in $\delta$, which is then set to the unit value, while 
$\eta$ is fixed by some optimization procedure.
The most popular prescription used in the literature is the PMS~\cite{Stevenson:1981vj}. The PMS  states that if a theory containing nonphysical parameters is an approximation of the correct theory, then varying the value of $\eta$ should not change the values of the physical quantities in the approximate theory.
Hence, 
for some physical quantity $\mathcal{O}_{\delta^k}$,  which is evaluated up to the order of $\delta^k$, the optimal value $\overline \eta$ is determined from
\begin{eqnarray}\label{PMS}
    \frac{\partial \mathcal{O}_{\delta^k}}{\partial \eta} {\bigg \vert}_{\overline \eta } = 0.
\end{eqnarray}

The first applications of the OPT method were done for systems at zero temperature~\cite{Seznec:1979ev,Stevenson:1981vj,Politzer:1981vc,Duncan:1988hw}. Later, it has also been applied at finite temperature at leading order~\cite{Okopinska:1986pd,Hajj:1987gk,Pinto:1999py,Kneur:2002kq}. Its applicability at finite temperature for the scalar $\lambda \phi^4$ theory in next-to-leading order was further tested in~\cite{Farias:2008fs}.
Here, we will borrow some of the main results from~\cite{Farias:2008fs} and refer the interested reader to that reference for more details. 

\begin{figure}
    \centering
    \includegraphics[width=1\linewidth]{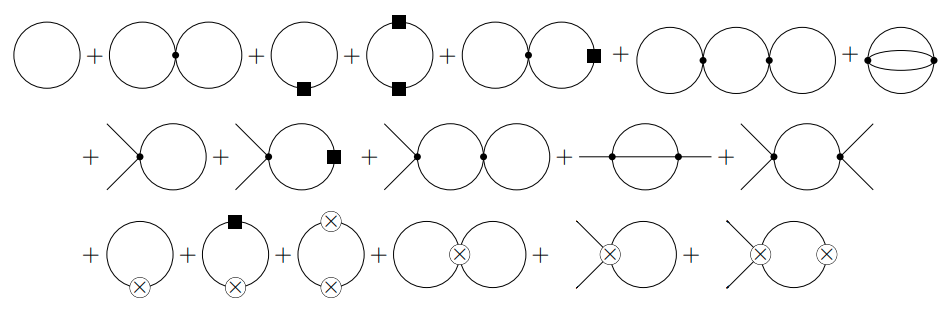}
    \caption{Feynman diagrams contributing to the effective potential for the $\lambda \phi^4$ model in the OPT case at NLO. The black circle 
represents a quartic vertex proportional to $\delta \lambda$, a black square is a (new) quadratic vertex proportional to $\delta \eta^2$ while the crossed circle represents renormalization counterterms. The external lines represent the background scalar field $\varphi$ in the
effective potential.}
    \label{fig1}
\end{figure}

In practice, when applying the OPT, one can  use the results obtained by standard perturbation theory up to some order $\lambda^k$ in the coupling
and then apply the changes
given by Eqs.~(\ref{Mmodopt}) and (\ref{Lmodopt}), while reexpanding the result up to the desired order $\delta^k$. In the present application, we are mainly concerned with
the finite temperature effective potential for the scalar quartic model at the order of $\lambda^2$. Applying Eqs.~(\ref{Mmodopt}) and (\ref{Lmodopt}) to the perturbative ${\cal O}(\lambda^2)$ standard  result and re-expanding yields
the corresponding OPT expression up to the order of $\delta^2$ (NLO), whose contributions are represented by the {}Feynman diagrams shown in {}Fig.~\ref{fig1}.

Within our procedure, the first step is to evaluate a relevant physical quantity, up to a given order, using the deformed theory described by the OPT Lagrangian density, Eq.(\ref{LrLct}). The effective potential can be obtained using the standard method in the literature~\cite{Dolan:1973qd,Jackiw:1974cv,Weinberg:1974hy,Buchbinder:1992rb} by shifting the scalar field around a background field $\phi \to \varphi + \phi'$, where $\langle \phi \rangle = \varphi$ and performing the functional integration over $\phi'$, after which we can use the substitutions (\ref{Mmodopt}) and (\ref{Lmodopt}) to obtain the corresponding OPT expression. Explicitly, the renormalized effective potential for the OPT at NLO at finite temperature is given by~\cite{Farias:2008fs}
\begin{equation}
V_{\rm OPT}= \Delta V_{\rm OPT}^{\delta^0} \delta^0 + \Delta V_{\rm OPT}^{\delta^1} \delta^1 + \Delta V_{\rm OPT}^{\delta^2} \delta^2 +
{\cal O}(\delta^3), 
\label{OPT2}
\end{equation}
where
\begin{eqnarray}
\Delta V_{\rm OPT}^{\delta^0} &=& \frac{1}{2}\left[ m^2+ \eta^2 \right]\varphi^2
\nonumber \\
&-&\frac{\Omega^{4}\left( 2 L_\Omega +3\right)}{8\left( 4\pi \right) ^{2}}-\frac{J_{0,\Omega} T^{4}}{2\left( 4\pi \right)^{2} } ,
\label{VOPT0}
\end{eqnarray}
\begin{eqnarray}
\Delta V_{\rm OPT}^{\delta^1} &=& - \frac{1}{2}\eta^2 \varphi^2+ \frac{\lambda}{4!} \varphi^4
\nonumber \\
&+& \frac{\hbar \eta ^{2}}{2\left( 4\pi \right) ^{2}}\left[ \left(L_\Omega +1\right) \Omega^{2}-J_{1,\Omega} T^{2}\right]  \nonumber \\
        &-& \frac{\varphi ^{2}}{2} \frac{ \lambda }{2(4\pi )^{2}}\left[\left( L_\Omega +1\right) \Omega^{2}-J_{1,\Omega} T^{2} \right] \nonumber\\
 &+& \frac{ \lambda }{8\left( 4\pi \right) ^{4}}\left[ \left( L_\Omega +1\right) \Omega^{2}-J_{1,\Omega} T^{2}\right] ^{2} 
 \nonumber\\
 &+& \mathcal{F}_{\delta},
\label{VOPT1}
\end{eqnarray}
and 
\begin{widetext}
\begin{eqnarray}
\Delta V_{\rm OPT}^{\delta^2} &=&   - \frac{ \eta ^{4}}{4\left( 4\pi \right) ^{2}}\left[L_\Omega +J_{2,\Omega} \right] - \frac{ \lambda }{4\left( 4\pi \right) ^{4}}\eta ^{2}\left[ L_\Omega +J_{2,\Omega}\right] \left[ \left( L_\Omega+1\right) \Omega^{2}-J_{1,\Omega} T^{2}\right]  \nonumber \\
&-&\frac{ \lambda ^{2}}{48\left( 4\pi \right) ^{6}}\left[\left( \left( 3L_\Omega+4\right) J_{1,\Omega}^{2} +J_{1,\Omega}^{2} J_{2,\Omega} +2K_{2,\Omega}+\frac{4}{3}K_{3,\Omega}\right) 3 T^{4}\right. \nonumber \\
&-& \left( 12L_\Omega^{2}+28L_\Omega - 12-\pi ^{2}-4C_{0}+6\left( L_\Omega +1\right) J_{2,\Omega} \right) J_{1,\Omega} \Omega^{2}T^{2} \nonumber \\
&+& \left. \left( 5 L_\Omega^{3}+17L_\Omega^{2} +\frac{41}{2} L_\Omega-23-\frac{23 \pi^2}{12} - \psi ^{\prime \prime }\left( 1\right) +C_{1} + 3\left( L_\Omega+1\right)^{2}J_{2,\Omega} \right) \Omega^{4}  \right]\nonumber \\
 &+& \frac{\varphi ^{2}}{2} \left[ - \frac{ \lambda^2 \Omega^2}{4(4\pi)^4} \left((L_\Omega+1)^2 + 1+\frac{\pi^2}{6}\right) + \frac{ \lambda \eta ^{2}}{2(4\pi) ^{2}}\left(L_\Omega +J_{2,\Omega}\right)
 +\frac{  \lambda^2}{4(4\pi)^4} \left(\left( L_\Omega +1\right) \Omega^{2}-J_{1,\Omega} T^{2} \right) \left(L_\Omega+ J_{2,\Omega}\right) \right. \nonumber \\
&-& \left. \frac{  \lambda ^{2}}{6(4\pi) ^{4}} \left[- 3\Omega^2 \left(L_{\Omega}^2 + 3 L_{\Omega} + C_2 \right) + 3 T^2 \left( L_\Omega J_{1,\Omega} +H_{2,\Omega} + H_{3,\Omega} \right)\right] \right] 
-\frac{\varphi ^{4}}{4!} \frac{3\lambda ^{2}}{2 (4\pi) ^{2}}\left[ L_\Omega+J_{2,\Omega}\right] + \mathcal{F}_{\delta^2}\;,
\label{VOPT2}
\end{eqnarray}
\end{widetext}
where we have  defined $L_\Omega = \ln(\mu^2/\Omega^2)$, with $\mu$ being the renormalization energy scale in the
$\overline{\rm MS}$ regularization scheme while
$C_0$, $C_1$ and $C_2$ are numerical constants given, respectively, by $C_0 = -9.8424$, $C_1 = 39.429$ and $C_2 = 3.33288$.
The quantities $J_{0,\Omega}$, $J_{1,\Omega}$, $J_{2,\Omega}$, $H_{2,\Omega}$, $H_{3,\Omega}$, $K_{2,\Omega}$ and $K_{3,\Omega}$ are all functions of $\Omega/T$ 
and their explicit expressions are given in Appendix~\ref{app:ThermalIntegrals}, while the expressions for the terms $\mathcal{F}_{\delta}$ and $\mathcal{F}_{\delta^2}$ appearing also in Eq.~(\ref{OPT2}) are provided in Appendix~\ref{app:physicalparameters}. 
In the expression for the effective potential, Eq.~(\ref{OPT2}), $\varphi$
denotes the constant background scalar field, while $m$ and $\lambda$ represent the renormalized mass and coupling, respectively. Note that  $m$ and $\lambda$ are implicit functions
of the renormalization scale $\mu$ and they are obtained from the loop expansion of the original theory, prior to the application of the OPT
prescription given by (\ref{Mmodopt}) and (\ref{Lmodopt}). 

The next step is the implementation of the RGI procedure to the   
the {\it original} loop expanded $V_{\rm eff}$ (see Appendix~\ref{app:originaleffectivepotential}). Once  the relevant  RGI $V_{\rm eff}$ is determined one may apply the OPT procedure in order to resum the perturbative series. In the next section we review these steps, starting by determining  the RGI $V_{\rm eff}$.

\section{Renormalization Group Improvement and the OPT Implementation}
\label{sec3}

In this section, we present a brief summary of the method that allows for the mitigation of scale dependence in an effective potential within perturbation theory in $\hbar$. In the context of the effective potential of the massive $\lambda \phi^4$ theory at zero temperature $V_{\rm eff}$ which explicit expression can be found in Ref.~\cite{Chung:1999xm}, the effective potential must satisfied the \textit{renormalizations group equation} (RGE), that can be written as~\cite{Kleinert:2001ax}
\begin{widetext}
\begin{eqnarray}\label{callansymanzik}
    \left[ \mu \frac{\partial}{\partial \mu} + \beta_\lambda \frac{\partial}{\partial \lambda} + \gamma_{m} m^2 \frac{\partial}{\partial m^2} + \beta_\Lambda \frac{\partial}{\partial \Lambda} - \gamma_\varphi\varphi \frac{\partial}{\partial \varphi}  \right] V_{\rm eff} (\varphi) = 0,
\end{eqnarray}
\end{widetext}
where the functions \(\beta_\lambda\), \(\gamma_{m}\), \(\beta_\Lambda\), and \(\gamma_\varphi\) are defined as
\begin{eqnarray}
    \beta_\lambda &\equiv& \mu  \frac{\partial \lambda}{\partial \mu}, \\
    \gamma_{m} &\equiv& \frac{\mu}{m^2}  \frac{\partial m^2}{\partial \mu}  ,\\
    \gamma_{\varphi} &\equiv& \mu  \frac{\partial}{\partial \mu} \ln Z_\varphi^{1/2} , \\
    \beta_{\Lambda} &\equiv& \mu  \frac{\partial \Lambda}{\partial \mu},
\end{eqnarray}
with $Z_\varphi$ representing the wave function renormalization term. Note that to implement the RG procedure, we also introduce the ``cosmological constant" $\Lambda$ \cite{Brown:1992db}. In the RGI context, $\Lambda$ is used to ensure the scale invariance of the effective potential \cite{Kastening:1991gv,Bando:1992np, Ford:1992mv}.
{}For the $\lambda \phi^4$ theory, the functions \(\beta_\lambda\), \(\gamma_{m}\), \(\beta_\Lambda\), and \(\gamma_\varphi\) are well known and are
defined, up to order $\hbar^3$, as~\cite{Kleinert:1991rg}
\begin{eqnarray}\label{funçoesdogrupolimitadas}
    \beta_\lambda &=& \beta_{0} \lambda^2 \hbar + \beta_{1} \lambda^3 \hbar^2 + \beta_{2} \lambda^4 \hbar^3 , \nonumber \\
    \gamma_{m} &=& \gamma_{m 0} \lambda \hbar + \gamma_{m 1} \lambda^2 \hbar^2+ \gamma_{m 2} \lambda^3 \hbar^3, \nonumber \\
    \gamma_\varphi &=& \gamma_{0} \lambda \hbar + \gamma_{1} \lambda^2 \hbar^2 + \gamma_{2} \lambda^3 \hbar^3, \nonumber \\
    \beta_\Lambda &=& m^4 \left(\beta_{\Lambda 0} \hbar + \beta_{\Lambda 1} \lambda \hbar^2 + \beta_{\Lambda 2} \lambda^2 \hbar^3 \right),
\end{eqnarray}
where
\begin{eqnarray}
    \beta_{0} &=& \frac{3}{(4\pi)^2} , \; \; \beta_{1} = - \frac{17}{3(4\pi)^4} ,\;\; 
   \beta_{2} = \frac{\frac{145}{8} + 12 \zeta(3)}{(4\pi)^6} , \nonumber \\
    \gamma_{m_0} &=& \frac{1}{(4\pi)^2} , \; \; \gamma_{m_1} = - \frac{5}{6(4\pi)^4}  , \;\; \gamma_{m_2} = \frac{7}{2(4\pi)^6} , \nonumber \\
    \gamma_{0} &=& 0  , \; \; \gamma_{1} = \frac{1}{12(4\pi)^4} , \;\; \gamma_{2} = -\frac{1}{16(4\pi)^6} , \nonumber \\
    \beta_{\Lambda_0} &=& \frac{1}{2(4\pi)^2} , \; \; \beta_{\Lambda_1} = 0 \; \; \text{, and} \;\; \beta_{\Lambda_2} =\frac{1}{16(4\pi)^6} .
\end{eqnarray}

An effective potential fully satisfying Eq.~(\ref{callansymanzik}) is invariant under the RGE. In practice, the literature presents some methods that provide ways to obtain an approximate solution to Eq.~(\ref{callansymanzik}) in a way that mitigates the scale dependence of the effective potential. For our purposes, the method introduced in Ref.~\cite{Kastening:1991gv} and later explored by the authors of Refs.~\cite{Bando:1992np, Ford:1992mv,Bando:1992wy, Ford:1994dt,Ford:1996hd} will prove to be the most useful. The improvement of the effective potential through the properties of the renormalization group has also been investigated in the context of finite temperature in Ref.~\cite{Nakkagawa:1996ju} at leading order in the loop expansion. The calculations for the scalar theory \(\lambda \phi^4\) at $T=0$  were later studied in Refs.~\cite{Chung:1999gi} and \cite{Chung:1999xm}
at two- and three-loop orders, respectively.

The method proposed in Ref.~\cite{Kastening:1991gv} involves solving the RGE through a set of ordinary differential equations order by order in perturbation theory. 
{}Following the original Refs.~\cite{Kastening:1991gv,Bando:1992np,Chung:1999gi,Chung:1999xm}, the RGI of the effective potential
begins by noticing that the solution of 
Eq.~(\ref{callansymanzik}) cannot vary if one considers a change in the energy scale from \(\mu\) to \(\bar{\mu}\) and the effective potential should remain unaffected through such a change of scale. {}Formally, this condition of invariance under a change of scale from \(\mu\) to \(\bar{\mu}\) can be expressed as 
\begin{eqnarray}\label{Vrgi}
    V_{\rm eff} (\mu, \lambda, m^2, \varphi, \Lambda) = V_{\rm eff} (\bar{\mu}, \bar{\lambda}, \bar{m}^2, \bar{\varphi}, \bar{\Lambda}).
\end{eqnarray}
{}Following the procedure in Refs.~\cite{Bando:1992np,Chung:1999gi}, to find a solution of
Eq.~(\ref{callansymanzik}), we introduce the reparametrization $\bar{\mu} \to t \bar{\mu}$, such that $\bar{\lambda}(\bar{\mu}) = \bar{\lambda}(t \bar{\mu})  , \;\; \bar{m}^2(\bar{\mu}) = \bar{m}^2(t \bar{\mu})  , \;\; \bar{\varphi}(\bar{\mu}) = \bar{\varphi}(t \bar{\mu}) \text{,} \;\; \bar{\Lambda}(\bar{\mu}) = \bar{\Lambda}(t \bar{\mu})$. Based on the requirement that the effective potential remains constant along the curves defined by this reparametrization, i.e.,
\begin{widetext}
\begin{eqnarray}\label{renspace}
\frac{dV_{\rm eff}}{dt} \equiv
   \left[ \frac{\partial\bar{\mu}}{\partial t} \frac{\partial}{\partial \bar{\mu}} + \frac{\partial \bar{\lambda}}{\partial t} \frac{\partial}{\partial \bar{\lambda}} + \frac{\partial \bar{m}^2}{\partial t} \frac{\partial}{\partial \bar{m}^2} + \frac{\partial \bar{\Lambda}}{\partial t} \frac{\partial}{\partial \bar{\Lambda}} +\frac{\partial \bar{\varphi}}{\partial t}\frac{\partial}{\partial \bar{\varphi}}  \right] V_{\rm eff}=0.
\end{eqnarray}
Comparing Eq.~(\ref{renspace}) with the RGE~(\ref{callansymanzik}), we identify the set of equations to 
be solved:
\begin{eqnarray}\label{setRGequations}
    \hbar \frac{d\bar{\mu}}{dt} = \bar{\mu}  \text{,} \; \; \; \hbar \frac{d \bar{\lambda}}{dt} = \bar{\beta}_\lambda \text{,} \; \; \; \hbar \frac{d\bar{m}^2}{dt} = \bar{\gamma}_{m} \bar{m}^2  \text{,} \; \; \; \hbar \frac{d\bar{\varphi}}{dt} = - \bar{\gamma}_\varphi \bar{\varphi}  \text{,} \; \; \; \hbar \frac{d\bar{\Lambda}}{dt} = \bar{\beta}_\Lambda,
\end{eqnarray}
\end{widetext}
where we have reintroduced the $\hbar$ symbol to keep track of the order in which these equations are solved (as a loop expansion, in powers of $\hbar$) for the parameters of the theory. Up to order $\hbar^2$, 
the solutions of the set of differential equations
in (\ref{setRGequations}) are
\begin{eqnarray}
    \bar{\lambda} &=& \bar{\lambda}_{0} + \bar{\lambda}_{1} \hbar + \bar{\lambda_{2}} \hbar^2, \label{expansionparameters1} \\
    \bar{m}^2 &=& \bar{m}_{0}^2 + \bar{m}_{1}^2 \hbar + \bar{m}_{2}^2 \hbar^2, \label{expansionparameters2} \\
    \bar{\phi} &=& \bar{\phi}_{0} + \bar{\phi}_{1} \hbar + \bar{\phi}_{2} \hbar^2, \label{expansionparameters3} \\
    \bar{\Lambda} &=& \bar{\Lambda}_{0} + \bar{\Lambda}_{1} \hbar + \bar{\Lambda}_{2} \hbar^2
    \label{expansionparameters4},
\end{eqnarray}
where the explicit expressions for each term in the above solutions can be found in Refs.~\cite{Chung:1999gi,Chung:1999xm}.
{}For completeness, we reproduce these solutions in Appendix~\ref{app:solutions} where they are explicitly presented in terms of  Eqs.~(\ref{lamb0}), (\ref{lamb1}), (\ref{lamb2})--(\ref{Lambdabar2}).

\subsection{Obtaining the effective potential in the RGI method}

The underlying physical motivation for the chosen RGI strategy can be summarized as follows. 
The RGI strategy adopted here is motivated by the 
requirement that physical observables should be insensitive to arbitrary renormalization 
scales, which arise as artifacts of the perturbative framework. In thermal field 
theory, this issue is further compounded by the presence of medium-induced effects, such as 
thermal screening, which are not efficiently captured by a naive perturbative expansion. The 
introduction of a variational parameter provides an effective way to encode these medium effects 
through a dynamically generated mass scale. By combining RG invariance with a variational 
optimization procedure, the VRG approach enforces a principle of minimal sensitivity with 
respect to both the renormalization scale and the variational parameter. This leads to a 
reorganization of the perturbative series around physically relevant scales, thereby improving 
the stability and reliability of the resulting thermodynamic quantities.

The RGI method applied to the effective potential (see, e.g., Ref.~\cite{Chung:1999gi}) starts by first substituting Eqs.~ (\ref{expansionparameters1})-(\ref{expansionparameters4}) into the original effective potential to ensure that Eq.~(\ref{Vrgi}) is satisfied. The effective potential is then reexpanded in powers of \(\hbar\) up to the appropriate order at which $V_{\rm eff}$ is being evaluated.  After reexpanding $V_{\rm eff}$, we then substitute in it the solutions
for each one of the terms appearing on the
right-hand side of Eqs.~ (\ref{expansionparameters1}) -  (\ref{expansionparameters4}), e.g., one uses the explicit solutions given by Eqs.~(\ref{lamb0}), (\ref{lamb1}), (\ref{lamb2})--(\ref{Lambdabar2}). The result still has a dependence on the renormalization scale and some suitable choice for $\mu$ must be considered. {}For this, we first note that the first relation in Eq. (\ref{setRGequations}), when using the boundary condition $\bar{\mu}(0) = \mu$, has the solution
\begin{eqnarray}\label{rgimubar}
    \bar{\mu}^2 = \mu^2 \exp( 2 t / \hbar).
\end{eqnarray}
We also note that the parameter \(t\) encodes how the change between the scales \(\mu\) and \(\bar{\mu}\) occurs. The main idea of the RGI method is to choose a value for \(t\) in such a way as to minimize the dependence on the scale. 
The scale dependence in the original effective potential arises from logarithmic terms. 
These scale-dependent logarithmic terms can vanish through an appropriate choice of \(\bar{\mu}\),
for example, by taking the argument of original logarithms to satisfy ~\cite{Bando:1992np}
\begin{eqnarray}\label{firstchoicemu}
    \frac{\bar{m}^2 (t) + \frac{1}{2} \bar{\lambda}(t) \bar{\phi}(t)^2}{\bar{\mu}^2(t)} = 1 ,
\end{eqnarray}
which must be solved order by order, since the parameters of the theory have a perturbative expansion. Although solving Eq.~(\ref{firstchoicemu}),  in general, represents a complex task, one can, nevertheless,  choose \(\bar{\mu}\) judiciously. As suggested in Ref.~\cite{Chung:1999xm}, by choosing the simple form
\begin{eqnarray}
    t = \frac{\hbar}{2} \ln \left(\frac{m^2 + \lambda \varphi^2/2}{\mu^2} \right),
\end{eqnarray}
one can reproduce Kastening's original results~\cite{Kastening:1991gv}, as shown in Ref.~\cite{Chung:1999gi}.

In the context of the finite-temperature effective potential, the same procedure can be applied. New terms arise due to the finite-temperature procedure; these terms also contain the parameters of the theory, which must undergo the appropriate modifications due to the RGI, for further details, see Ref.~\cite{Nakkagawa:1996ju}.

Next, we describe the procedure outlined above for the RGI of the effective potential in the context of the OPT method at finite temperatures.

\subsection{Obtaining the effective potential in the VRG method}

We now present a consistent implementation of the VRG method for the finite-temperature effective potential derived within 
the OPT. The strategy is to apply OPT directly to the RGI finite-temperature effective potential, 
thereby combining variational resummation with a systematic reduction of renormalization-scale dependence.

The construction proceeds as follows. {}First, we consider the loop expanded finite-temperature effective potential. The RGI procedure described above is then applied to this effective potential. At this stage, 
the RG equations are solved without fixing the optimized scale $\bar{\mu}$. This procedure leads to the RGI effective potential
as described above.

Next, the OPT deformation, defined by Eqs.~(\ref{Mmodopt}) and (\ref{Lmodopt}),  is implemented in the RGI effective potential. When expanding the effective potential in powers of the bookkeeping parameter $\delta$, the quantity $\xi = 1 - \beta_0 \lambda t$
is left unexpanded. This prescription ensures that the $\delta$ expansion does not spoil the logarithmic resummation achieved by the RGI method, in close analogy with the treatment of the $\hbar$ expansion in Refs.~\cite{Chung:1999gi,Chung:1999xm}. 

The optimum scale $\bar{\mu}$ is next fixed according to the OPT order under consideration. 
{}For instance, at high temperatures and at first order in $\delta$, all logarithmic terms can be cast in the form 
$\ln(\alpha T/\bar{\mu})$, where $\alpha = 4\pi/e^{\gamma_E}$. This allows us to choose, for example, 
$\bar{\mu} = \alpha T$, and adopt the same prescription at order $\delta^2$. 
The resulting VRG effective potentials at orders $\delta$ and $\delta^2$ are given in Eqs.~(\ref{VRG1}) and (\ref{VRG2}), 
respectively, with explicit expressions collected in Appendix~\ref{app:vrgdelta2}.
The final step is to obtain the arbitrary mass parameter $\eta$, which is then fixed by applying the PMS condition Eq.~(\ref{PMS}) to the VRG effective potential.

\section{Results}
\label{sec4}

Let us now analyze how well the VRG performs in order to attenuate the scale dependence of the OPT effective potential by mainly comparing predictions for the symmetric ($m^2\ge 0$) and the nonsymmetric 
($m^2 <0$) cases. Concerning the first case,  we will also make a comparison with the results provided by other nonperturbative methods commonly used in the literature, e.g., the results obtained from the two-particle irreducible approach (2PI), from the functional renormalization group (FRG) and from the RGOPT method. This will help us gauge how our approach performs compared to these other methods. In addition, we also examine whether the VRG potential preserves the universality class of the \(\lambda \phi^4\) theory\footnote{Recall that the $Z_2$ symmetric scalar field model~(\ref{Lagr})  belongs to the same universality class of the Ising model for~$d\leq4$, with the model exhibiting a second-order phase transition at the critical point~\cite{Zinn-Justin:1989rgp,Kleinert:2001ax,Kenna:1993fp}.}.

To analyze the dependence of the effective potential on the renormalization scale, we start by considering the scale dependence of the renormalized parameters $m(\mu)^2$, $\lambda(\mu)$, and $\varphi(\mu)$. As already specified in the previous section, at the first order for both OPT and VRG, we consider the running of the renormalized parameters up to ${\cal O}(\hbar^2)$ and which are determined by the equations:
\begin{eqnarray}
    \mu \lambda^{\prime}(\mu) &=& 3 \frac{\lambda(\mu)^2}{(4\pi)^2} - \frac{17 \lambda(\mu)^3}{3(4\pi)^4} , \label{lambdamu1} \\
    \mu \frac{m^{\prime}(\mu)}{m(\mu)} &=& \frac{\lambda(\mu)}{2(4\pi)^2}  , \label{mmu1} \\
    \mu \frac{\varphi^{\prime}(\mu)}{\varphi(\mu)} &=& - \frac{\lambda(\mu)^2}{12(4\pi)^4},
    \label{phimu1}
\end{eqnarray}
where these solutions are used in Eq.~(\ref{OPT2}), at order $\delta$, in the case of OPT, and in Eq.~(\ref{VRG1}) for VRG.
In the second order for both OPT and VRG, we consider the running of the parameters up to ${\cal O}(\hbar^3)$, which are now given by
\begin{eqnarray}
    \mu \lambda^{\prime}(\mu) &=& 3 \frac{\lambda(\mu)^2}{(4\pi)^2} - \frac{17 \lambda(\mu)^3}{3(4\pi)^4} 
    \nonumber \\
    &+& \frac{\lambda(\mu)^4}{(4\pi)^6} \left[\frac{145}{8} + 12 \zeta(3)\right], \label{lambdamu2}\\
    \mu \frac{m^{\prime}(\mu)}{m(\mu)} &=& \frac{\lambda(\mu)}{2(4\pi)^2} + \frac{7\lambda(\mu)^2}{2(4\pi)^6}  , \label{mmu2}\\
    \mu \frac{\varphi^{\prime}(\mu)}{\varphi(\mu)} &=& - \frac{\lambda(\mu)^2}{12(4\pi)^4} -\frac{\lambda(\mu)^3}{16(4\pi)^6}. \label{phimu2}
\end{eqnarray}
where we now substitute these solutions in Eq.~(\ref{OPT2}), up to order $\delta^2$, in the case of OPT, and in Eq.~(\ref{VRG2}) for VRG.
The boundary conditions used to solve the set of Eqs.~(\ref{lambdamu1})-(\ref{phimu2}) are: $\lambda(\mu_0) = \lambda_0$, $m(\mu_0) = m_0$, and  $\varphi(\mu_0) = \varphi_0$ where $\mu_0$ is a reference scale to be defined below. As is common in the literature, to check the stability of the computed quantities with the scale $\mu$, we will vary it around the so-called ``central" value $\mu=2 \pi T$ within the usual range $\mu \in [\pi T, 4 \pi T]$. As a reference scale, we can then choose $\mu_0 = 2\pi T_0$ where, for simplicity, the reference temperature is set to $T_0 = m_0 /(2 \pi)$ so that $\mu_0 = m_0$. This choice will allow us to easily express all physical quantities in units of $m_0$ (or likewise, in terms of the reference scale $\mu_0$). 

\subsection{Symmetric phase}

\begin{center}
\begin{figure}[!htb]
\subfigure[]{\includegraphics[width=8.5cm]{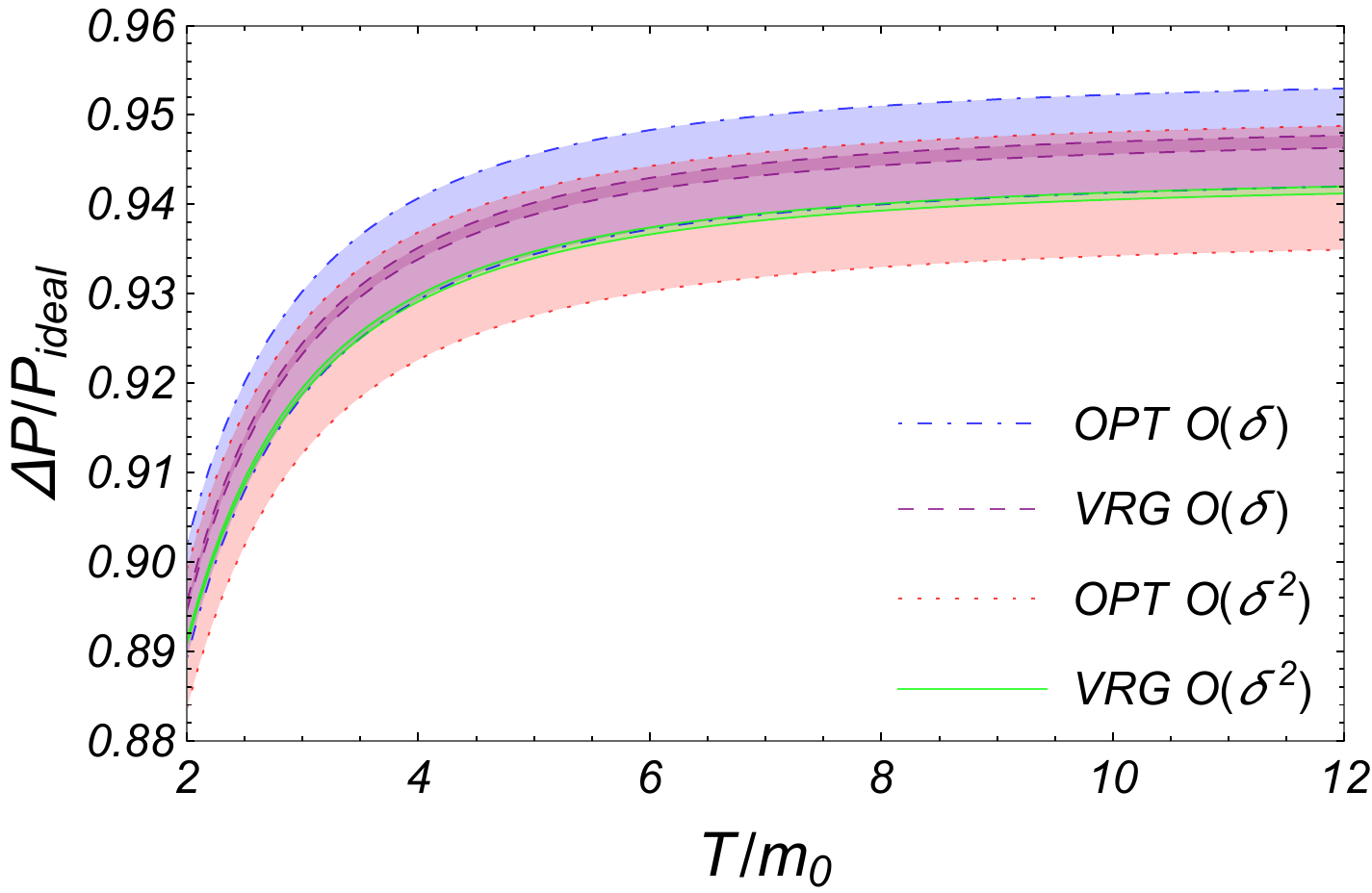}}
\subfigure[]{\includegraphics[width=8.5cm]{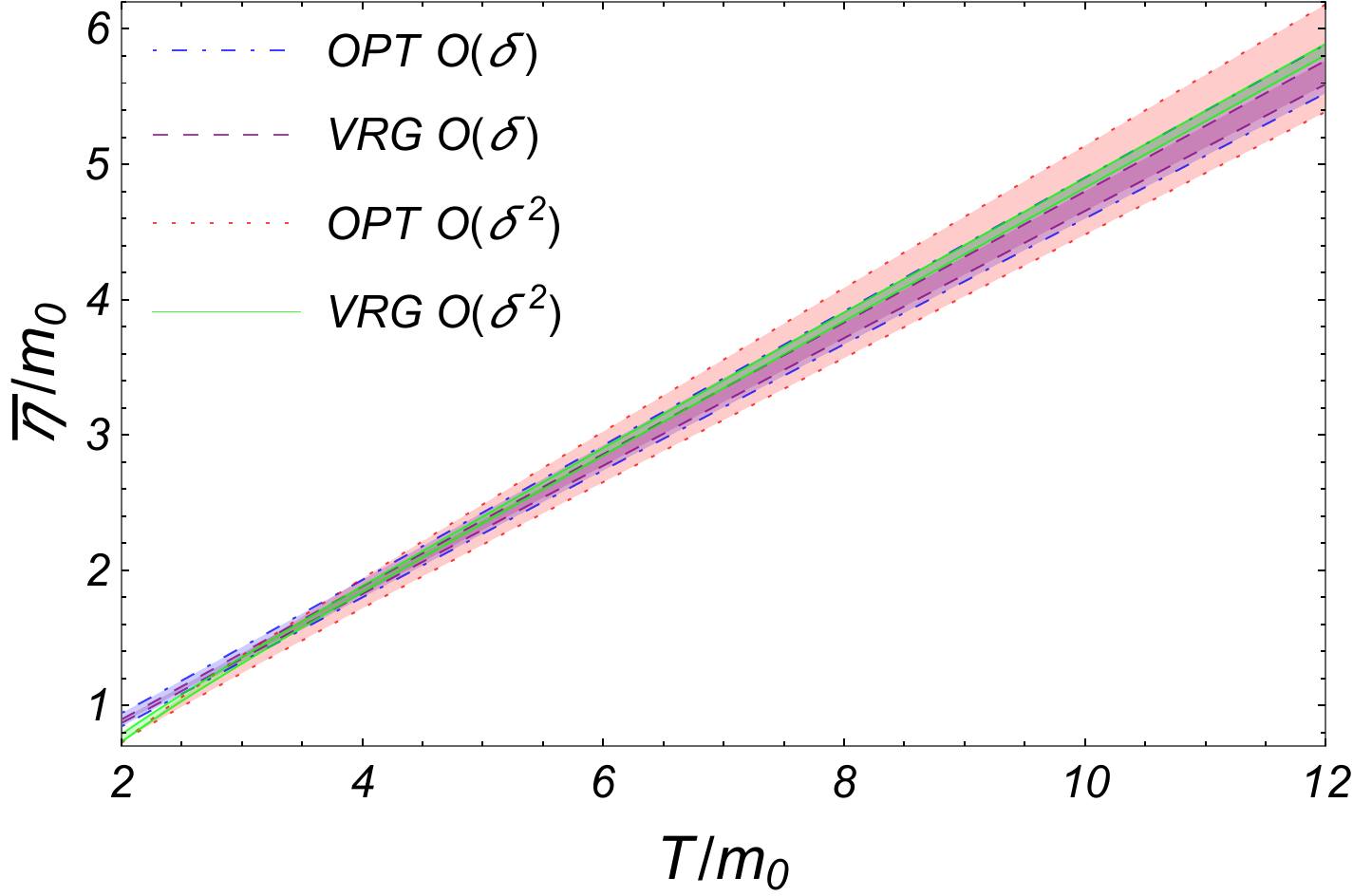}}
\caption{The pressure subtracted by the constant vacuum term, $\Delta P = P-P_{\rm vacuum}$, normalized by the ideal gas result (panel a) and the optimal mass parameter \(\overline \eta\) (panel b) as functions of $T/m_0$ for the OPT and VRG methods at orders $\delta$ and $\delta^2$. In both cases, the coupling value is fixed at the representative value \(\lambda_0 = 12.25\) and the scale dependence range is given by \(\pi T \le \mu \le 4\pi T\). In the OPT up to $\delta$ and $\delta^2$ order, and VRG up to $\delta^2$, the upper curve in the bands corresponds to the value $\mu=4\pi T$, while the lower curve in the bands corresponds to $\mu=\pi T$. Otherwise, the VRG up to $\delta$ order gets inverted. This pattern is repeated in the other figures.}
  \label{fig2}
\end{figure}
\end{center}

The symmetric phase is characterized by a non-negative quadratic field term in the potential, which leads to a vanishing vacuum expectation value, $\langle \phi \rangle\equiv\varphi=0$. Let us start by analyzing, within the different schemes, the pressure and which in the symmetric phase is then defined as $P = - V_{\rm eff}(\varphi=0,T)$.  With this aim, it is convenient to normalize $P$ by the ideal gas value, which, for a free scalar field theory, is given by  
\begin{eqnarray}  
    P_{\text{ideal}} = \frac{\pi^2 T^4}{90}. 
\end{eqnarray}  
In {}Fig.~\ref{fig2}, we compare the OPT and VRG results for the pressure subtracted by the constant vacuum term, $\Delta P = P-P_{\rm vacuum}$,
(panel a) and for the optimal variational PMS parameter $\bar \eta$ as functions of temperature (panel b).
The results indicate that the VRG exhibits a much milder scale dependence than the OPT. We also find that in both cases, the different orders ($\delta$ and $\delta^2$) show a very good convergence for $\Delta P$, with it stabilizing between \(0.92 \leq \Delta P/P_{\text{ideal}} \leq 0.95\) in the temperature range considered. In {}Fig.~\ref{fig2}(b) indicates that the optimal $\overline \eta$ provided by the VRG prescription is less sensitive to scale variations than the result from the OPT. This suggests that the new approach encodes part of the improvement brought about by the renormalization group also directly on the optimal value $\bar \eta$.

\begin{center}
\begin{figure}[!htb]
\subfigure[]{\includegraphics[width=8.5cm]{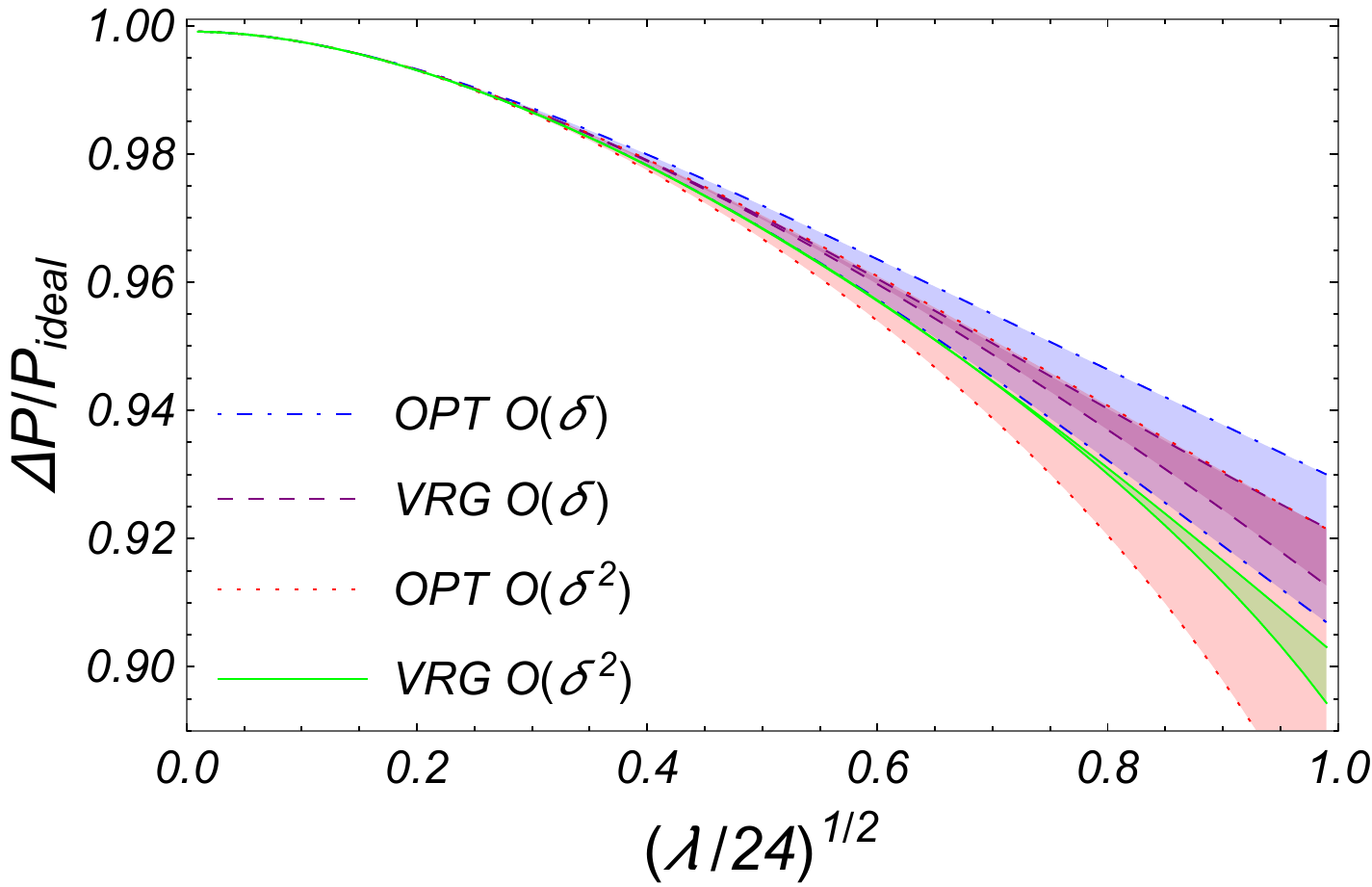}}
\subfigure[]{\includegraphics[width=8.5cm]{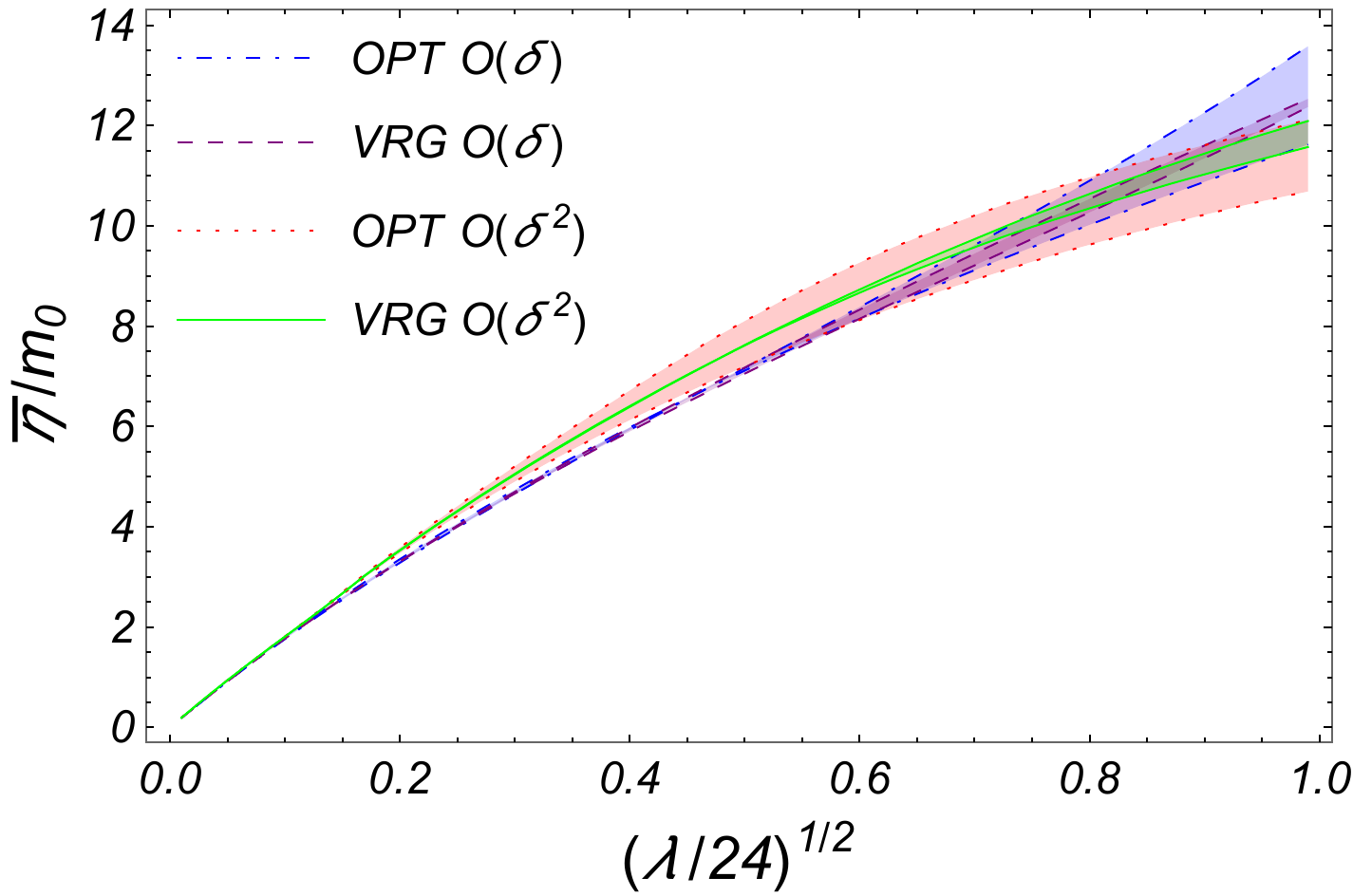}}
\caption{Similar to {}Fig.~\ref{fig2} but  showing the normalized pressure (panel a) and the optimal mass parameter \(\overline \eta\) (panel b) as a function of the renormalized coupling, with the temperature fixed at \(T= 20 m_0\).}
  \label{fig3}
\end{figure}
\end{center}

In {}Fig.~\ref{fig3} we now show the results for the pressure (panel a) and for the optimal PMS mass parameter $\bar \eta$ (panel b) as a function of the renormalized coupling, while keeping the temperature fixed.  We observe that the VRG predictions remain close to the center of the band generated by the OPT, while significantly reducing the scale dependence at both orders. This is the same qualitative behavior observed in {}Fig.~\ref{fig2}. {}From the value \(\lambda \gtrsim 1\), or equivalently $(\lambda/24)^{1/2} \gtrsim 0.2$, the bands showing the scale dependence increase considerably in the case of the OPT, while for the VRG the increase is less dramatic.

\begin{center}
\begin{figure*}[!htb]
\subfigure[]{\includegraphics[width=8.5cm]{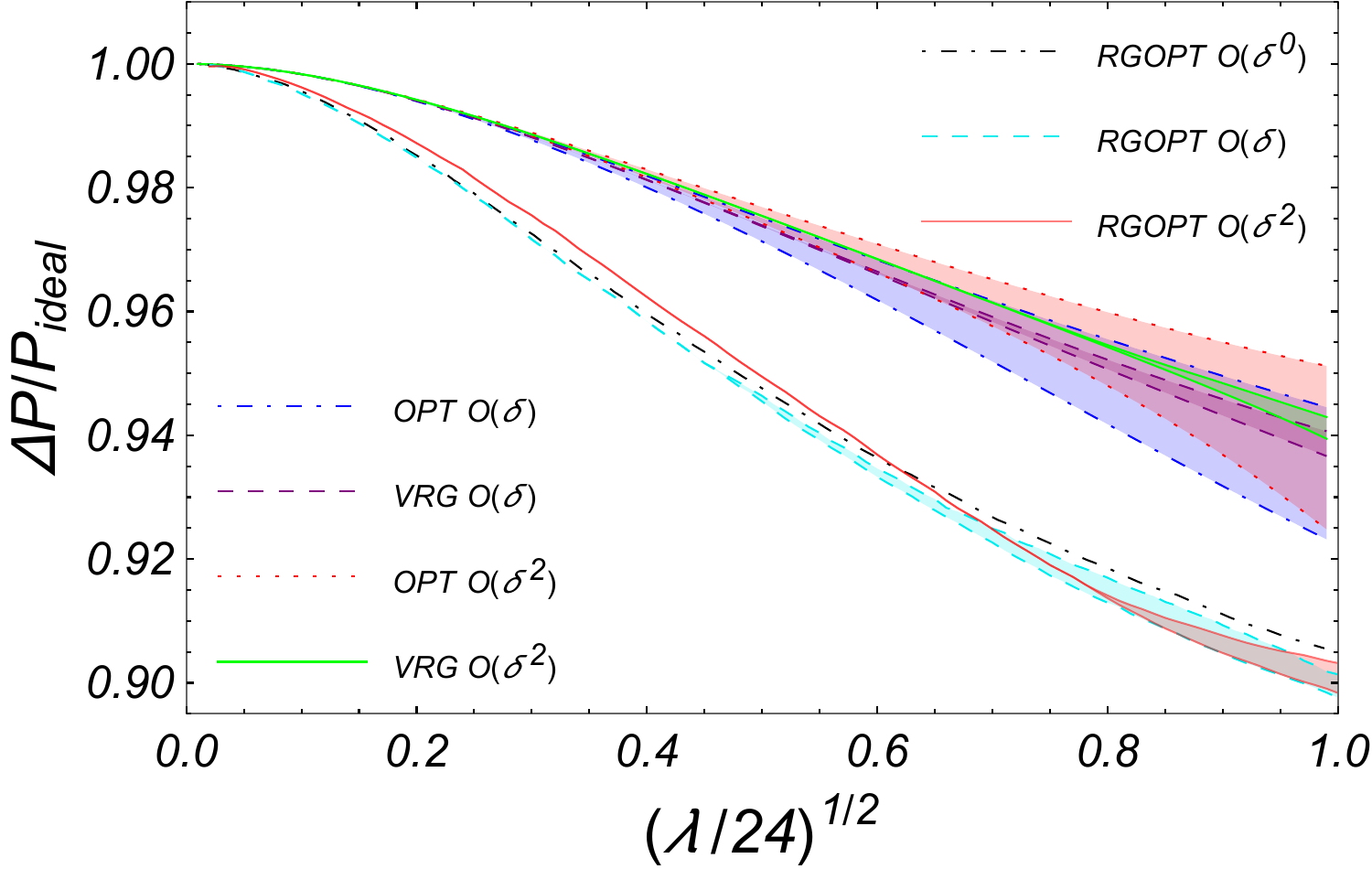}}
\subfigure[]{\includegraphics[width=8.5cm]{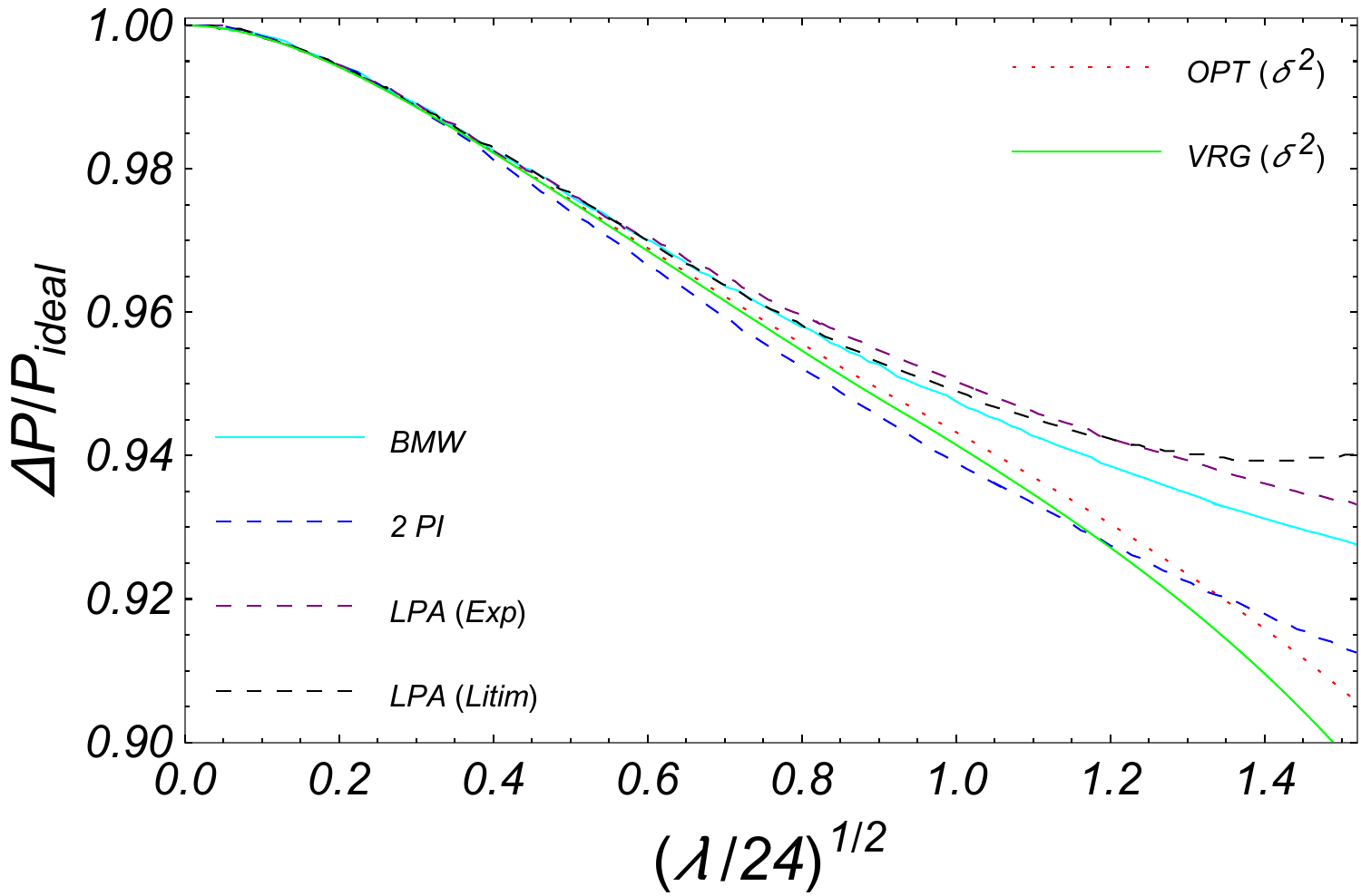}}
\caption{The pressure subtracted by the constant vacuum term, $\Delta P = P-P_{\rm vacuum}$, normalized by the ideal gas result, as a function of the coupling, for the (symmetric) massless case. In panel (a), the OPT, VRG, and RGOPT results are compared for \(\pi T \le \mu \le 4\pi T\). In panel (b), the OPT, VRG, FRG (with implementations LPA and BMW---see text), and 2PI predictions are compared for the central scale, $\mu = 2\pi T$. The renormalized coupling is taken at the central value  for the scale, $\lambda\equiv\lambda(2\pi T)$ and the temperature is fixed at $T = \mu_0$.}
  \label{fig4}
\end{figure*}
\end{center}

In addition to the direct comparison between the two techniques shown in {}Figs.~\ref{fig2} and \ref{fig3}, the comparison can also be included with other methods found in the literature. Many of these alternative approximations present results for the pressure that are very much similar to the one shown in {}Fig. \ref{fig3}(a) for example.

While the strong reduction of the renormalization-scale dependence observed in 
the VRG results provides an important consistency check, it should not be interpreted 
by itself as a proof of improved physical accuracy. In the present framework, this 
stabilization arises from the combined optimization with respect to both the RG scale 
and the variational parameter, which enforces a form of local insensitivity to unphysical 
parameters. This procedure effectively reorganizes the perturbative expansion and 
typically correlates with improved convergence properties. However, part of the reduction 
in $\mu$ dependence is also a direct consequence of the variational constraint and, therefore, 
does not strictly guarantee proximity to the exact result. A definitive assessment of the 
accuracy of the method ultimately requires comparison with nonperturbative benchmarks, which is
here done in {}Fig.~\ref{fig4}, with results from other available nonperturbative methods found in the literature.
To  compare, in {}Fig.~\ref{fig4}(a),   our results with the predictions of other methods, we  only consider the massless limit since this is the  case analyzed by the other authors. In {}Fig.~\ref{fig4}(a) we  compare our results for the OPT and  VRG with those produced by the RGOPT. The latter  results (up to order-$\delta^2$) were originally obtained in Ref. \cite{Fernandez:2021sgr}. Note that this reference also presents the predictions   from the screened perturbation theory (SPT), which turn out to be very similar to the ones generated by the standard OPT. Therefore, to make the comparison less clumsy, in {}Fig.~\ref{fig4}(a) we compare only OPT, VRG and RGOPT. 

{}For the same massless (unbroken symmetry) case,  the literature also contains results generated by alternative methods such as the functional renormalization group (FRG), which involves a possible truncation of the potential and a choice of regulator. The local potential approximation (LPA) is commonly used to solve the FRG flow equation, and the regulators employed in this approximation were the exponential regulator and the Litim regulator (see Ref.~\cite{Blaizot:2006rj} for details).  Reference~\cite{Blaizot:2010ut} also includes the Blaizot, M\'endez-Galain and Wschebor (BMW) approximation which is  based on the FRG   with the aim to improve the LPA one. In addition to these FRG-based methods, results obtained with the 2PI resummation were also presented in Refs.~\cite{Blaizot:2006rj,Blaizot:2010ut}. To make a clear comparison of our results with all these other methods, in {}Fig.~\ref{fig4}(b) we consider only the OPT and VRG cases at order $\delta^2$. Also, since those other methods have presented results only at the central scale value $\mu=2\pi T$ (without showing the 
scale-dependent bands), we do the same here for the OPT and VRG.
As shown in Fig.~\ref{fig4}(b), the VRG (and similarly the OPT) remains close to the 2PI results up to 
$(\lambda/24)^{1/2} \sim 1.2$ (corresponding to $\lambda \sim 34.6$), beyond which the VRG 
exhibits an increased suppression of the pressure relative to the other methods. 
On the other hand, the RGOPT, shown in Fig.~\ref{fig4}(a), also indicates a similar, though more pronounced, suppression of 
the pressure relative to the ideal gas case.

\begin{center}
\begin{figure}[!htb]
\includegraphics[width=8.5cm]{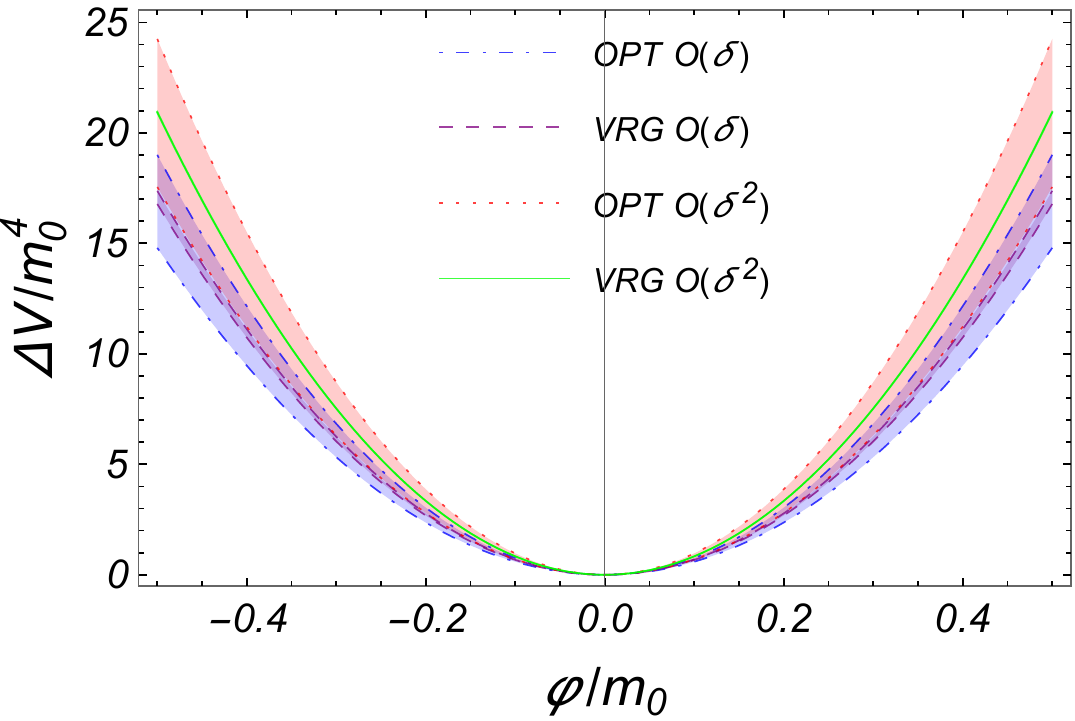}
\caption{Subtracted effective potential, $\Delta V = V_{\rm eff}(\varphi,T) - V_{\rm eff}(\varphi=0,T)$, in  units of mass as a function of the field for  fixed coupling \(\lambda_0 =12.25\) and temperature ($T= 20 m_0$). The scale dependence range is given by \(\pi T \le \mu \le 4\pi T\). The figure shows the results obtained with the OPT and VRG at perturbative orders \(\delta\) and \(\delta^2\).}
  \label{fig5}
\end{figure}
\end{center}

Let us now examine the subtracted effective potential, $\Delta V= V_{\rm eff}(\varphi,T) - V_{\rm eff}(\varphi=0,T)$, as a function of the background field, $\varphi$. The results are shown in
{}Fig.~\ref{fig5} for the choices $T = 20 m_0$ and   \(\lambda_0 = 12.25\). It can be observed in this figure that, as the field increases, the scale dependence becomes more pronounced in the case of OPT at both orders, whereas in the case of VRG, also at both orders, the effective potential is not so  sensitive to scale variation as the field grows.

\subsection{Broken phase}

Let us now consider the results for the broken symmetry phase, where investigations related to the phase transitions can be performed.
This will also allow us to gauge the scale dependence on the critical temperature for symmetry restoration.
In the broken phase, the field acquires a temperature-independent nonvanishing vacuum expectation value, $\sigma(0)$, already at the classical (tree) level. Then, considering the temperature-dependent higher-order contributions to the effective potential, one can analyze the thermal behavior of the order parameter, $\sigma(T)$, which characterizes the possible phase transition patterns. The critical temperature ($T_c$) associated with the restoration of symmetry is determined by the condition $\sigma(T_c)=0$. At order $\delta$, the authors of Ref.~\cite{Farias:2008fs} found that the OPT free energy leads to a first-order phase transition failing to respect the universality class of the $\lambda \phi^4$ model. The very same result is also found here with the VRG. However, at order $\delta^2$, the OPT correctly predicts a second-order phase transition, as shown in Ref.~\cite{Farias:2008fs}.  As we shall see, at the same perturbative order, the VRG also correctly predicts a second-order phase transition. {}For this reason, we will now focus on only the order-$\delta^2$ results when analyzing the broken phase.

As usual, the order parameter $\sigma(T)$ can be determined by minimizing the thermal effective potential, which here is done for both the OPT and VRG ones,
\begin{eqnarray}\label{VEVef}
    \left. \frac{d V_{\rm eff}(\varphi, T)}{d\varphi} \right|_{\varphi = \sigma(T)} = 0.
\end{eqnarray}  
Another quantity of interest is the (temperature-dependent) curvature of the effective potential at the origin\footnote{Even though it is common to call Eq.~(\ref{massatermica}) a thermal mass, it should not be confused with the true temperature dependent effective mass, defined through the on-shell pole of the scalar field propagator at finite temperature. At order $\delta$ both quantities are the same, but at second order, since the effective potential only includes off-shell contributions, they are not equivalent. },
\begin{eqnarray}\label{massatermica}
    m_T^2 = \left. \frac{d^2 V_{\rm eff}(\varphi)}{d\varphi^2} \right|_{\varphi = 0}.
\end{eqnarray}
Note that the above derivative is taken around the origin with the aim of investigating the phase transition, that is, the sign change of the curvature term of the potential. We also emphasize  that Eq.~(\ref{massatermica}), at the level of the approximations considered here, yields exactly the same critical temperature result as Eq.~(\ref{VEVef})\footnote{
Note that this equality holds within the present level of approximation used in the VRG framework.
At higher orders, small deviations may arise due to residual truncation effects.
In other situations (e.g., first-order phase transitions), where linear and 
cubic terms in the field may be present (either at tree level or generated 
by quantum/thermal corrections), this equality would no longer hold.}.

\begin{center}
\begin{figure*}[!htb]
\subfigure[]{\includegraphics[width=7.5cm]{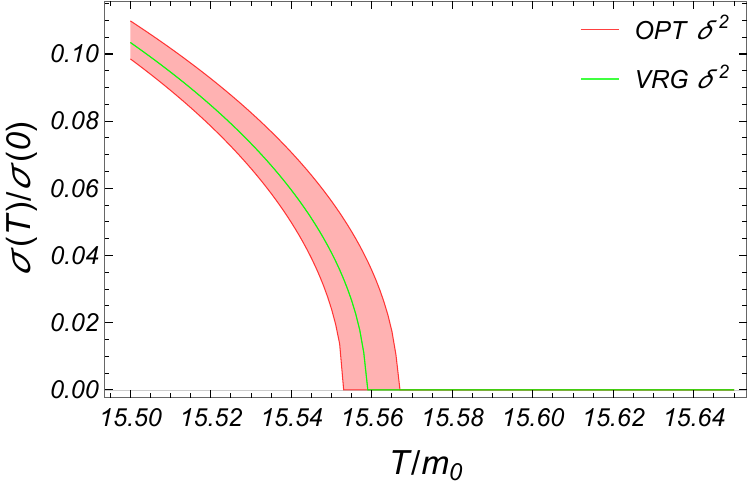}}
\subfigure[]{\includegraphics[width=8.2cm]{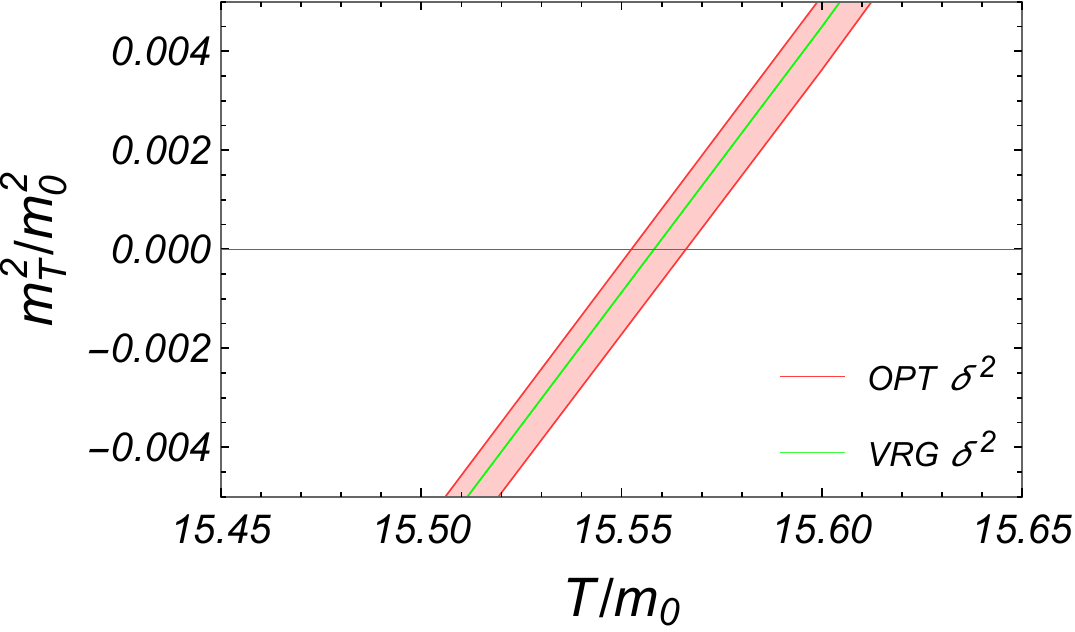}}
\caption{The temperature dependent expectation value normalized by the tree-level vacuum value $\sigma(0)$ (panel a)
and the effective potential curvature (panel b) as a function of the temperature. The scale dependence range is given by \(\pi T \le \mu \le 4\pi T\) and the coupling is fixed at \(\lambda_0 = 0.1\).  In both cases, the results for OPT and VRG are at order \(\delta^2\). {}For the OPT, the curve on the right corresponds to $\mu=\pi T$, while the curve on the left corresponds to $\mu=4\pi T$. Similar for VRG, though the band due to the variation of the scale is barely visible in the plots.}
  \label{fig6}
\end{figure*}
\end{center}

\begin{center}
\begin{figure*}[!htb]
\subfigure[]{\includegraphics[width=8.2cm]{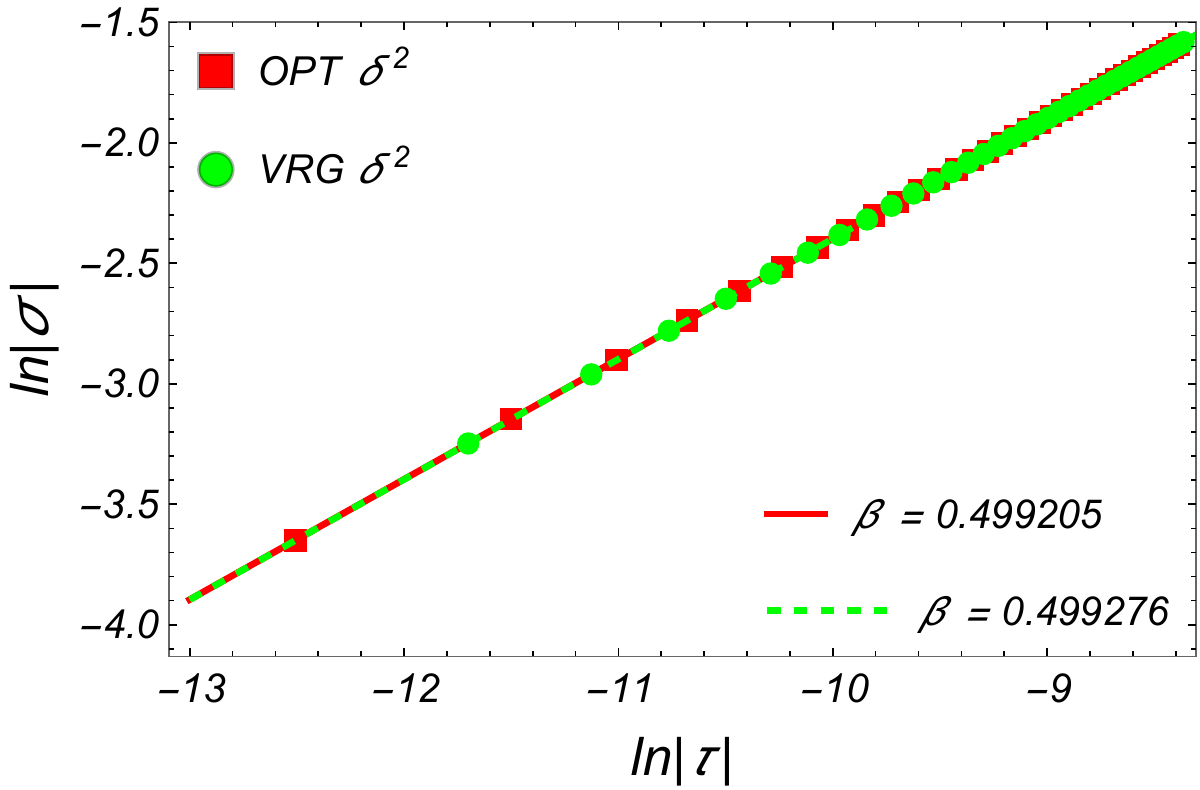}}
\subfigure[]{\includegraphics[width=8.2cm]{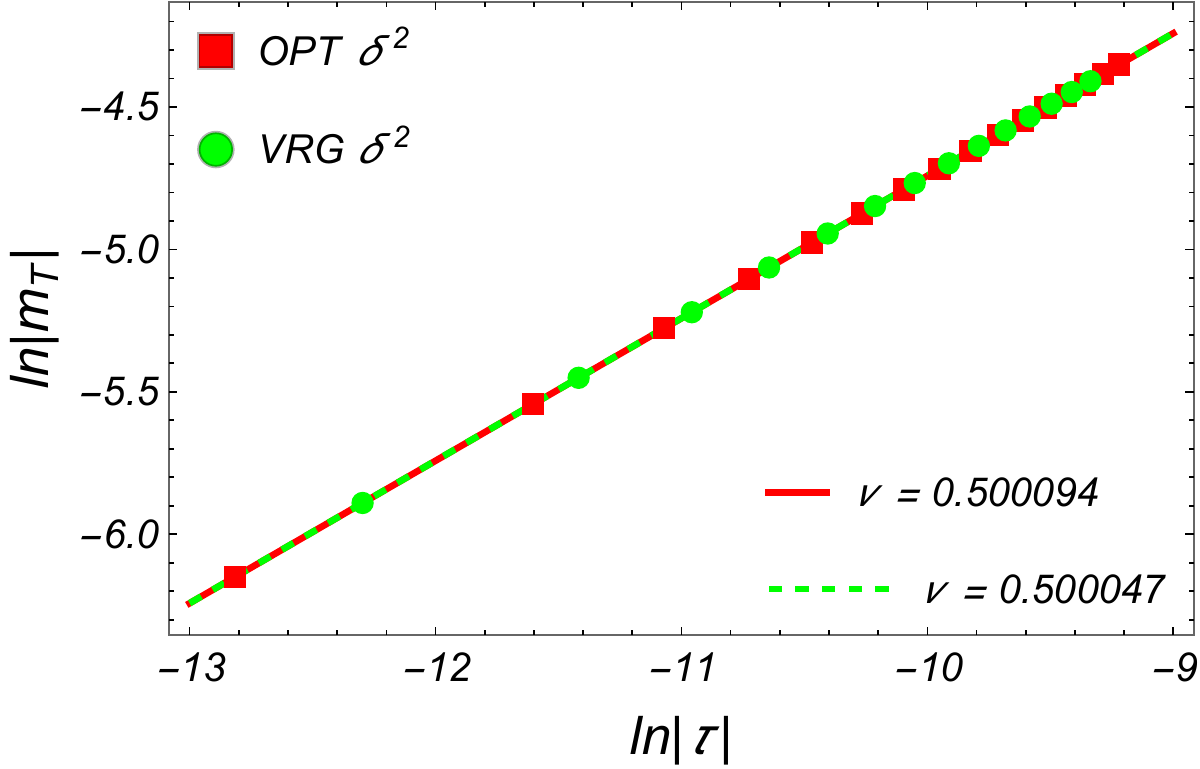}}
\caption{The results for $\ln |\sigma|$ (panel a) and 
for $\ln |m_T|$ (panel b) as a function of $\ln |\tau|$.
In each case, the dots are the numerical results. The corresponding values for the critical exponents in each case, obtained by the fittings, are also indicated. The coupling is fixed at $\lambda_0 = 0.1$ and $\mu = \pi T$. }
  \label{fig7}
\end{figure*}
\end{center}

In {}Fig.~\ref{fig6} we show the temperature dependence for $\sigma(T)$ (panel a) and for $m_T^2$ (panel b) for both OPT and VRG at order $\delta^2$. To facilitate visualization of the scale dependence of both quantities, here we have considered the coupling as fixed at the value $\lambda_0 = 0.1$, while $\mu$ is again varied within the range \(\pi T \le \mu \le 4\pi T\). Although it may appear that there is no scale variation in the VRG, this is merely a misleading impression caused by the image, since a rather mild scale variation was observed at the numerical level. This figure is important because, in addition to illustrating the continuous variation of the vacuum expectation value of the field, it also provides an assessment of how $T_c$ responds to scale variations, as we shall discuss in what follows. 
Note that the temperature-dependent curvature $m_T^2$ of the effective potential, which is another way to determine the critical temperature, expresses the variation of the mass values and a change in its sign marks the phase transition. Let us point out that the coupling was chosen at \(\lambda_0 = 0.1\) for two main reasons: the first is that weak couplings require higher critical temperatures, favoring the high-temperature analysis used in this work; the second reason is that a weak coupling allows us to perform a direct comparison with results obtained in the literature, such as in perturbative calculations~\cite{Dolan:1973qd,Weinberg:1974hy} and nonperturbative calculations~\cite{Banerjee:1991fu,Pinto:1999py,Farias:2008fs}.

{}Figure~\ref{fig6} also shows that $\sigma(T)$ and $m_T^2$ approach the critical point in a way that is similar to that exhibited by the Ising model in three spatial dimensions in the context of statistical mechanics~\cite{Parisi:1988nd}, and can be specified by the critical exponents defined as follows:
\begin{eqnarray}
\nu = \lim_{\tau \to 0} \frac{\ln |m_T|}{\ln|\tau|},
\label{nu}
\\
\beta = \lim_{\tau \to 0} \frac{\ln \sigma(\tau)}{\ln|\tau|},
\label{betac}
\end{eqnarray}
where, in the usual prescription, $\tau$ denotes the reduced temperature,
$\tau=(T-T_c)/T_c$ and $T_c$ is the critical temperature.
In {}Fig.~\ref{fig7}, a linear fitting is performed for both the thermal expectation value (panel a) and the curvature of the potential (panel b). In panel (a), we start from a temperature below the critical temperature and gradually increase it in successive steps until it approaches the critical temperature, with each point corresponding to a temperature increment of \(\Delta T/m_0 = 0.0001\). Panel (b) basically displays the same procedure but starting from a temperature above the critical value.  Note that both results approach the critical exponent values $\nu=1/2$ and $\beta=1/2$, which are still the values predicted by the mean-field approximation~\cite{Kleinert:2001ax}.
A similar situation has also been shown to occur in the two-loop $\Phi$-derivable  approximation~\cite{Marko:2012wc}, which also found the critical exponents to coincide with those in the mean-field approximation. In that reference, this was attributed to the order of the approximation used not being enough to produce nonanalyticities in the effective potential. We believe that a similar issue also occurs here, despite the fact that OPT and VRG at order $\delta^2$ correctly predict a second-order phase transition.

We note that, due to the high-temperature approximation considered here, weak couplings ($\lambda_0 \ll 1$) make the scale dependence range quite tight. However, a numerical analysis of the critical temperature values can be obtained and shall prove to be useful when further analyzing the scale dependence of a physical observable such as $T_c$. In Table~\ref{tab:tempcrit}, we present the $T_c$ values for $\lambda_0=0.1, 0.5$ and $1.0$, as well as the quantity $\Delta T_c = [T_c (\pi T) - T_c (4\pi T)]/T_c (\pi T)$, which represents the percentage variation in the critical temperatures computed at the extrema of the scale variation that we consider.

\begin{widetext}
\begin{center}
\begin{table}[!htb]
 \caption{Sample results from OPT and VRG for the critical temperature and its percentual difference at the extrema of the interval for the scale dependence, \(\pi T \le \mu \le 4\pi T\).}
 \label{tab:tempcrit}
    \begin{tabular}{c  c  c  c  c}
    \hline
  $\lambda_0$ & $[T_c^{OPT} (\pi T),\; T_c^{OPT} (4\pi T)]$ & $\Delta T_c^{OPT}$ (\%) & $[T_c^{VRG} (\pi T),\; T_c^{VRG} (4\pi T)]$ & $\Delta T_c^{VRG}$( \%) \\
    \hline
    $ 0.1$ & $[15.566558 , \;15.552963]$ & $0.087333$ & $[15.558629, \; 15.558627]$ & $0.000010$ \\
    $ 0.5$ & $[7.00635 ,\; 6.97562]$ & $0.438617$ & $[6.988491, \; 6.988452]$ & $0.000558$ \\
    $ 1.0$ & $[4.964792,\; 4.920350]$ & $0.895167$ & $[4.939094, \;4.938934]$ & $0.003237$\\
    \hline
    \end{tabular}
\end{table}
\end{center}
\end{widetext}


\begin{center}
\begin{figure}[!htb]
\includegraphics[width=8.2cm]{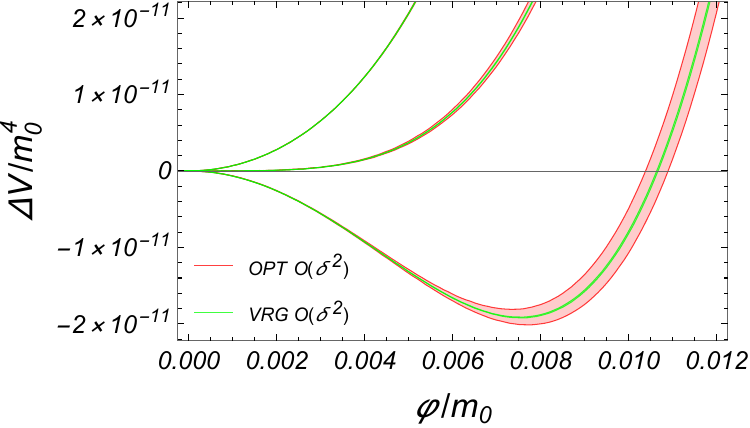}
\caption{Subtracted effective potential, $\Delta V = V_{\rm eff}(\varphi,T) - V_{\rm eff}(\varphi=0,T)$, as a function of $\varphi$. The results for OPT and VRG are at order \(\delta^2\). The coupling is fixed at the values $\lambda_0 = 1$. The scale dependence range is such that \(\pi T \le \mu \le 4\pi T\). {}From bottom to top, the regions correspond to the temperature fixed at \(T =T_c - \Delta T\), \(T = T_c\), and \(T = T_c + \Delta T\), respectively, with \(\Delta T/m_0 = 5 \times 10^{-6}\).}
  \label{fig8}
\end{figure}
\end{center}

Let us now check
the dependence of the effective potential in the OPT and VRG cases as a function of the background field just as we have done in the symmetric case (see {}Fig.~\ref{fig5}). This is shown in {}Fig.~\ref{fig8}. We choose $\lambda_0=1$ as a representative value for the coupling constant. The values of the temperature have been taken below and above the critical value, which shows well that the phase transition is second-order, as also confirmed with the results shown in {}Fig.~\ref{fig6}.
{}Furthermore, the results displayed in {}Fig.~\ref{fig8} indicate
the efficiency of the VRG method in suppressing the scale dependence compared to the results for the OPT.
An important aspect elucidated by these results is the difference in scale at the minimum of the potential, a feature that had already been demonstrated in {}Fig.~\ref{fig6}(a).

\section{Conclusions}\label{conclusions}

In this work, we present an alternative resummation method that aims to improve the effective potential obtained with the traditional OPT by imposing RG conditions. The proposed new prescription, called here the VRG, combines the RGI method, originally prescribed in Refs.~\cite{Chung:1999gi,Chung:1999xm} with the OPT procedure. The OPT is directly applied to the RGI finite-temperature effective potential, which then combines the variational resummation with a systematic reduction of renormalization-scale dependence.

This approach encapsulates a consistent way to merge these two tools and the results presented here for the $\lambda \phi^4$ theory are promising, showing a significant reduction in the scale dependence in both the symmetric and broken phases
of the theory. A major advantage of the prescription proposed in this work is that no modification of the standard OPT framework is required apart from the incorporation of the properties of the RGI procedure. Since VRG improves the scale dependence of the standard OPT, we believe that it can also be useful in improving the scale of other thermal resummation methods, such as SPT and HTLpt, which are plagued by high scale dependence issues.

In the broken phase, the VRG has proven to be fully consistent for what we expect for the phase transition pattern of the theory. In particular, the predictions for the order parameter and the critical temperature have proven to be very stable against scale variations. We have observed that the VRG respects the same universality class as the $\lambda \phi^4$ theory, predicting  a second-order phase transition, as expected. In this application, it was not possible to access the critical temperature for higher couplings because of the use of high-temperature approximations. This may be remediated in the future by accounting for the full momentum integration of the loop terms, bypassing the high-temperature approximations. This would require to perform a more comprehensive numerical analysis to better understand the broken phase for a wide range of $\lambda$ and temperature values. We leave this more involved numerical analysis for a future work. 

\begin{acknowledgements}

The authors thank Jean-Lo\"ic Kneur for discussions concerning the method considered here. 
The authors acknowledge financial support of the Coordena\c{c}\~ao de
Aperfei\c{c}oamento de Pessoal de N\'{\i}vel Superior (CAPES) -
{}Finance Code 001. R.O.R. is also partially supported by research
grants from Conselho Nacional de Desenvolvimento Cient\'{\i}fico e
Tecnol\'ogico (CNPq), Grant No. 307286/2021-5, and from
{}Funda\c{c}\~ao Carlos Chagas Filho de Amparo \`a Pesquisa do Estado
do Rio de Janeiro (FAPERJ), Grant No. E-26/201.150/2021. M.B.P. is partially supported by Conselho
Nacional de Desenvolvimento Cient\'{\i}fico e Tecnol\'{o}gico (CNPq),
Process  No.  307261/2021-2  and 403016/2024-0.

\end{acknowledgements}

\appendix

\section{Thermal Integrals}\label{app:ThermalIntegrals}

The thermal functions $J_n(a)$ appearing in Eq.~(\ref{OPT2}), with $a=\Omega/T$, are defined as
\begin{equation}
J_n(a ) \equiv \frac{4 \Gamma\left(\frac{1}{2}\right)}
{\Gamma \left( \frac{5}{2} -n\right) }
\int_0^{\infty} dx \frac{x^{4-2n}}{\sqrt{x^2+ a^2}}
\frac{1}{e^{\sqrt{x^2+a^2}}-1},
\label{Jn}
\end{equation}
which satisfies the identity
\begin{equation}
    J_{n+1} (a) = -\frac{1}{2a} \frac{\partial J_n(a)}{\partial a}.
\end{equation}
When $a\ll 1$, we have the high-temperature expansion for the $J_n(a)$ functions as given by~\cite{Farias:2008fs,Dolan:1973qd,Bellac:2011kqa,Gardim:2007ta}
\begin{eqnarray}
J_{0}(a) &=&
\frac{8\pi }{3}a^{3}+a^{4}\left( \ln \left( \frac{a }{4\pi }\right) +\gamma_E -
\frac{3}{4}\right)-\frac{4\pi ^{2}}{3}a^{2}+\frac{16}{45}\pi^{4} \nonumber \\
&+&128\sum_{n=1}^\infty\frac{\left(-1\right)^{n}\left( 2n-1\right) !!\zeta
\left( 2n+1\right) a^{\left( 2n+4\right) }}{
32\left( n+2\right) !2^{n+1}\left( 2\pi \right)^{2n}},
\label{J0}
\end{eqnarray}
\begin{eqnarray}
J_{1}(a) &=&
-4\pi a -2a^{2}\left[ \ln
\left( \frac{a }{4\pi }\right) +\gamma_E -\frac{1}{2}\right] +\frac{
4\pi ^{2}}{3}\nonumber \\
&-&16\sum_{n=1}^\infty\left( \frac{\left( -1\right)^{n}\left( 2n-1\right)
!!\zeta \left( 2n+1\right) a^{\left( 2n+2\right) }}
{4n!2^{n+1}\left( n+1\right) \left( 2\pi \right)^{2n}}
\right),\nonumber \\
\label{J1}
\end{eqnarray}
\noindent
and
\begin{eqnarray}
J_{2}(a) &=&
\frac{2\pi }{a}+2\ln \left( \frac{a}{4\pi }
\right) +2\gamma_E \nonumber \\
&+&4\left[ \sum_{n=1}^\infty\frac{\left(-1\right)^{n}\left( 2n-1\right) !!
\zeta\left( 2n+1\right) a^{2n}}{n!2^{n+1}\left( 2\pi
\right)^{2n}}\right],
\label{J2}
\end{eqnarray}
where $\gamma_E = 0.57721$ is the Euler-Mascheroni constant. 

The function $H_{2}$ appearing in Eq.~(\ref{OPT2}) is given by~\cite{Andersen:2000yj}
\begin{eqnarray}\label{H2}
    H_2(a) &=&\left(2-\frac{ \pi}{ \sqrt{3}}\right) J_1(a),
\end{eqnarray}
while the $H_{3}$ function in the high-temperature approximation is~\cite{Parwani:1991gq}
\begin{eqnarray}
    H_3(a) &\simeq& -\frac{(4\pi)^2}{12}\left[\ln\left(a^2\right) + 5.3025 \right] \;.
\end{eqnarray}

The functions $K_{2}$ and $K_3$ in Eq.~(\ref{OPT2}) are, respectively, given by\cite{Andersen:2000zn,Fernandez:2021sgr}
\begin{eqnarray}
    K_2(a) &\simeq&  \frac{(4\pi)^4}{72}\left(\ln a +\frac{1}{2} +\frac{\zeta'(-1)}{\zeta(-1)} \right)\nonumber \\
    &-& 372.65\,a \left(\ln a +1.4658\right), 
    \label{K2hT}
\end{eqnarray}
and 
\begin{eqnarray}
    K_3(a) &\simeq&  \frac{(4\pi)^4}{48}\left( -\frac{7}{15} +\frac{\zeta'(-1)}{\zeta(-1)} -\frac{\zeta'(-3)}{\zeta(-3)} \right) \nonumber \\
    &+& 1600.0\, a \left(\ln a +1.3045\right),
     \label{K3hT}
\end{eqnarray}
where $\zeta(x)$ is the Riemann zeta function.

\section{Renormalized Parameters}\label{app:physicalparameters}

In the context of the usual perturbation theory, the renormalized parameters are determined by the renormalization conditions (e.g. by the pole of the propagator for the mass and by the amplitude $2 \to 2$ scattering for the quartic coupling constant)~\cite{Fradkin:2021zbi}. These parameters are also known as physical parameters~\cite{Andersen:2000zn}. Accordingly, the bare mass and coupling are related to the renormalized ones through
\begin{eqnarray}
    m_b^2 &= & m^2 +\frac{ \lambda m^2}{2(4\pi)^2} \left[\ln \left(\frac{\mu^2}{m^2}\right)+1\right] + \frac{\lambda^2 m^2}{(4\pi)^4} \left[- \frac{11}{48} \right. \nonumber \\
    &+& \left. \frac{1}{2} \ln^2\left(\frac{\mu^2}{m^2}\right)+\frac{1}{3} \ln \left(\frac{\mu^2}{m^2}\right) \right] + \mathcal{O}(\lambda^3)\;, \label{massafisica}\\
    \lambda_b &= &\lambda  + \frac{ \lambda^2}{(4\pi)^2} \left[\frac{3}{2} \ln\left(\frac{\mu^2}{m^2}\right) + 1\right]+ \mathcal{O}(\lambda^3) . \label{acoplamentofisico}
\end{eqnarray}
The effective potential for the OPT at order $\delta^2$, Eq.~(\ref{OPT2}), is obtained by first performing the above replacements in the perturbative effective potential with bare parameters
(see, e.g., Ref.~\cite{Farias:2008fs}) and then applying the OPT procedure given by Eqs.~(\ref{Mmodopt}) and (\ref{Lmodopt})
and expanding in $\delta$ to the desired order.
To order $\delta^2$, this procedure then gives origin Eq.~(\ref{OPT2}), with the two terms in that equation,
${\cal F}_{\delta}$ and ${\cal F}_{\delta^2}$, given, respectively, by
\begin{eqnarray}
   {\cal F}_{\delta} &=&\frac{\lambda  \varphi ^2 \Omega^2}{4 (4\pi)^2}\left[L_{\Omega}+1\right] + \frac{ \lambda  \Omega^2}{4 (4\pi)^4} \left[L_{\Omega}+1\right] \nonumber \\
    &\times& \left[\Omega^2 \left(L_{\Omega}+1\right)-T^2 J_{1,\Omega}\right],
\end{eqnarray}
and
\begin{eqnarray}
   {\cal F}_{\delta^2} &=& \frac{ \eta^2 \lambda }{4 (4\pi)^4} \left[J_{2,\Omega} (L_{\Omega}+1) \Omega ^2
   \right. 
   \nonumber \\
   &-&  \left. J_{1,\Omega} T^2 (2 J_{2,\Omega}+L_{\Omega})\right]  \nonumber \\
    &-& \left. \frac{ \lambda \varphi^2 \eta ^2}{2 (4\pi)^2}L_{\Omega} + \frac{\lambda ^2 \varphi ^4}{48 (4\pi)^2}(3 L_{\Omega}+2) + \frac{\hbar^3\lambda ^2}{48 (4\pi)^6} \left[\Omega^2 \right. \right. \nonumber \\
    &\times& \left. \left. J_{1,\Omega} T^2 \left(4 C_1-6 J_{2,\Omega} (L_{\Omega}+1) -12 L_{\Omega}^2-28 L_{\Omega}\right. \right. \right. \nonumber \\
    &+& \left. \left. \left. \pi ^2+12\right)\right] +\frac{\hbar^3\lambda ^2}{48 (4\pi)^6} \left[\Omega^4 (L_{\Omega}+1)\left(-4 C_1-\pi ^2 \right. \right. \right. \nonumber \\
    &+& \left. \left. \left.3 J_{2,\Omega} (L_{\Omega}+1)+10 L_{\Omega}-18\right)+3 J_{1,\Omega}^2 (3 L_{\Omega}+2) \right. \right. \nonumber \\
    &\times& \left.  T^4 ] +\frac{ \lambda ^2 \varphi ^2}{48 (4\pi)^4} \left[6 J_{1,\Omega} (3 L_{\Omega}+2) T^2-\Omega ^2 \left(4 C_1 \right. \right. \right. \nonumber \\
    &+& \left. \left. \left. 24 C_2 +6 J_{2,\Omega} L_{\Omega}+6 J_{2,\Omega}
\right.\right.\right.
\nonumber \\
&+&  \left. \left. 12 L_{\Omega}^2+32 L_{\Omega}-\pi^2 - 
12\right)\right] . 
\end{eqnarray}

\section{The perturbative effective potential of the massive $\lambda \phi^4$ theory in $\hbar$ order}\label{app:originaleffectivepotential}

In this appendix, we present the effective potential of the massive $\lambda \phi^4$ theory at finite temperature, organized in orders of $\hbar$. The zero-temperature case was used at the beginning of Sec~\ref{sec3} for the review of RGI, whose result is found in Ref.~\cite{Chung:1999xm}
\begin{eqnarray}\label{originalT0}
V_{\rm eff}&=& \Delta V_{\rm eff}^{\hbar^0} \hbar^0 + \Delta V_{\rm eff}^{\hbar^1} \hbar^1 + \Delta V_{\rm eff}^{\hbar^2} \hbar^2 \nonumber \\
&+& \Delta V_{\rm eff}^{\hbar^3} \hbar^3 + {\cal O}(\hbar^4),
\end{eqnarray}
where
\begin{eqnarray}
\Delta V_{\rm eff}^{\hbar^0} &=& \frac{1}{2}m^2 \varphi^2 + \frac{\lambda}{4!} \varphi^4 + \Lambda ,
\end{eqnarray}
\begin{eqnarray}
\Delta V_{\rm eff}^{\hbar^1} &=& - \frac{m_H^{4}\left( 2 L_{m_H} +3\right)}{8\left( 4\pi \right) ^{2}}-\frac{J_{0,m_H} T^{4}}{2\left( 4\pi \right)^{2} },
\end{eqnarray}
\begin{eqnarray}
\Delta V_{\rm eff}^{\hbar^2} &=& \frac{ \lambda }{8\left( 4\pi \right) ^{4}}\left[ \left( L_{m_H} +1\right) m_H^{2}-J_{1,m_H} T^{2}\right] ^{2} \nonumber \\
        &-& \frac{\lambda^2\varphi^2}{12 (4\pi)^4} \left[- 3m_H^2 \left(L_{m_H}^2 + 3 L_{m_H} + C_2 \right) \right. \nonumber \\
        &+& \left. 3 T^2 \left(L_{m_H} J_{1,m_H} +H_{2,m_H} + H_{3,m_H} \right)\right] \nonumber \\
        &-& \frac{\lambda ^2 \varphi ^2 m_H^2}{8 (4\pi)^4} \left[(L_{m_H}+1)^2 + 1+\frac{\pi^2}{6}\right]\;,
\end{eqnarray}
and 
\begin{widetext}
\begin{eqnarray}
\Delta V_{\rm eff}^{\hbar^3} &=& \frac{ \lambda ^{2}}{48\left( 4\pi \right) ^{6}}\left[\left( \left( 3L_{m_H}+4\right) J_{1,m_H}^{2} +J_{1,m_H}^{2} J_{2,m_H} +2K_{2,m_H}+\frac{4}{3}K_{3,m_H}\right) 3 T^{4}\right. \nonumber \\
&-& \left( 12L_{m_H}^{2}+28L_{m_H} - 12-\pi ^{2}-4C_{0}+6\left( L_{m_H} +1\right) J_{2,m_H} \right) J_{1,m_H} m_H^{2}T^{2} \nonumber \\
&+& \left. \left( 5 L_{m_H}^{3}+17L_{m_H}^{2} +\frac{41}{2} L_{m_H}-23-\frac{23 \pi^2}{12} - \psi ^{\prime \prime }\left( 1\right) +C_{1} + 3\left( L_{m_H}+1\right)^{2}J_{2,m_H} \right) m_H^{4}  \right] + {\cal O}(\lambda^3)\;,
\end{eqnarray}
\end{widetext}
where we have  defined $L_{m_H} = \ln(\mu^2/m_H^2)$ and $m_H^2 = m^2 + \frac{1}{2} \lambda \varphi^2$. Terms beyond ${\cal O}(\lambda^2)$ will not be necessary for the purposes of this work. This potential contains all orders in perturbation theory in $\lambda$. The quantities $J_{0,m_H}$, $J_{1,m_H}$, $J_{2,m_H}$, $H_{2,m_H}$, $H_{3,m_H}$, $K_{2,m_H}$ and $K_{3,m_H}$ are all functions of $m_H/T$ and their explicit expressions are given in Appendix~\ref{app:ThermalIntegrals}.

\section{Solutions for the RGI functions}\label{app:solutions}

Starting with the first differential equation in Eq.~(\ref{setRGequations}),
\begin{eqnarray}
    \hbar \frac{d\bar{\mu}}{dt} = \bar{\mu},
\label{dmudt}
\end{eqnarray}
and choosing the initial condition at \( t = 0 \) as \( \bar{\mu}(0) = \mu \), the solution of Eq.~(\ref{dmudt}) is given by
\begin{eqnarray}
    t = \frac{\hbar}{2} \ln \left(\frac{\bar{\mu}^2}{\mu^2} \right).
\end{eqnarray}

Considering now the equation for \(\bar{\lambda}\) in Eq.~(\ref{setRGequations})and following~\cite{Chung:1999gi,Chung:1999xm}, at one-loop order we have that
\begin{eqnarray}
    \bar{\lambda}_0^{\prime}(t)- \beta_0 \bar{\lambda}_0(t)^2 = 0,
\end{eqnarray}
whose solution, with initial conditions at \(t=0\) given by \(\bar{\lambda}_0(0) = \lambda(\mu)\), is found to be
\begin{eqnarray}\label{lambdabar0}
    \bar{\lambda}_{0}&=&\frac{\lambda}{\xi},
\label{lamb0}
\end{eqnarray}
where we have defined $\xi= 1 - \beta_0  \lambda t$. 
Going to \(\hbar^2\)-order, the equation for the scale-dependent coupling is
\begin{eqnarray}
    &&\hbar^2 \left[-2 \beta_0 \bar{\lambda}_0(t) \bar{\lambda}_1(t)-\beta_1 \bar{\lambda}_0(t)^3+\bar{\lambda}_1'(t)\right] \nonumber \\
    &+& \hbar \left[\bar{\lambda}_0'(t)-\beta_0 \bar{\lambda}_0(t)^2\right] = 0.
\label{lambda1part}
\end{eqnarray}
Using the solution from Eq.~(\ref{lambdabar0}), the Eq.~(\ref{lambda1part}) simplifies to
\begin{eqnarray}
    \hbar^2 \left[\frac{\beta_0 \lambda ^3}{(\beta_0 \lambda  t-1)^3}+\frac{2 \beta_0 \lambda  \bar{\lambda}_1(t)}{\beta_0 \lambda  t-1}+\bar{\lambda}_1'(t)\right] = 0,
\end{eqnarray}
and its solution, using the initial condition \(\bar{\lambda}_1(0) = 0\), is given by
\begin{eqnarray}\label{lambdabar1}
    \bar{\lambda}_{ 1}&=&-\frac{\beta_1\lambda^2}{\beta_0 \xi^2}\ln \xi.
\label{lamb1}
\end{eqnarray}
Similarly, at order \(\hbar^3\) we find
\begin{eqnarray}
    &-&\frac{\lambda ^4 \left(\beta_1^2 \ln ^2(1-\beta_0 \lambda  t)-3 \beta_1^2 \ln (1-\beta_0 \lambda  t)+\beta_0 \beta_2\right)}{\beta_0 (\beta_0 \lambda  t-1)^4}\nonumber \\
    &+&\frac{2 \beta_0 \lambda  \bar{\lambda}_2(t)}{\beta_0 \lambda  t-1}+\bar{\lambda}_2'(t) = 0,
\label{lambda2part}
\end{eqnarray}
where we have used the previous solutions for $\bar \lambda_0$
and $\bar \lambda_1$.  Considering the initial condition \(\bar{\lambda}_2(0) = 0\), we obtain the solution for Eq.~(\ref{lambda2part}) as given by
\begin{eqnarray}\label{lambdabar2}
   \bar{\lambda}_{2}&=&\frac{\lambda^3}{\xi^2}\left[\left(-\frac{\beta_1^2}{\beta_0^2}
+\frac{\beta_2}{\beta_0}\right)[\xi^{-1}-1]
-\frac{\beta_1^2}{\beta_0^2}\frac{\ln \xi}{ \xi} \right. \nonumber \\
&+& \left. {\beta_1^2}{\beta_0^2}\frac{\ln^2 \xi}{\xi}\right].
\label{lamb2}
\end{eqnarray}

The solutions for the scale-dependent mass at orders $\hbar^0$, $\hbar$ and $\hbar^2$ are found similarly (using the boundary conditions \(\bar{m}_0^2(0) = m^2(\mu)\) and \(\bar{m}_1^2(0) = \bar{m}_2^2(0) = 0\)) and are, respectively, given by 
\begin{eqnarray}\label{mbar0}
    \bar{m}_0^2 &=& \frac{m^2}{\xi^{\gamma_{m0}/\beta_0}},
\end{eqnarray}
\begin{eqnarray}\label{mbar1}
\bar{m}_1^2 &=& \frac{\lambda m^2}{\xi^{\gamma_{m0}/\beta_0}}\left[\left(
-\frac{\beta_1\gamma_{m0}}{\beta_0^2}+\frac{\gamma_{m1}}{\beta_0}\right)[\xi^{-1}-1] \right. \nonumber \\
&-& \left. \frac{\beta_1\gamma_{m0}}{\beta_0^2}\frac{\ln \xi}{\xi}\right],
\end{eqnarray}
and
\begin{widetext}
\begin{eqnarray}\label{mbar2}
\bar{m}_2^2 &=& \frac{\lambda^2 m^2}{\xi^{\gamma_{m0}/\beta_0}}\left[\left(\frac{\beta_1^2\gamma_{m0}}{\beta_0^3}
-\frac{\beta_2\gamma_{m0}}{\beta_0^2}-\frac{\beta_1^2\gamma_{m0}^2}{\beta_0^4} + \frac{2\beta_1\gamma_{m0}\gamma_{m1}}{\beta_0^3} - \frac{\gamma_{m1}^2}{\beta_0^2}\right)[\xi^{-1}-1] \right.
\nonumber \\
&+& \left.\left( 
-\frac{\beta_1^2\gamma_{m0}}{2\beta_0^3} + \frac{\beta_2\gamma_{m0}}{2\beta_0^2}
+\frac{\beta_1^2\gamma_{m0}^2}{2\beta_0^4}-\frac{\beta_1\gamma_{m1}}{2\beta_0^2}- \frac{\beta_1\gamma_{m0}\gamma_{m1}}{\beta_0^3} +\frac{\gamma_{m1}^2}{2\beta_0^2}
+\frac{\gamma_{m2}}{2\beta_0}\right)[\xi^{-2}-1] + \frac{\ln \xi}{\xi}\left(-\frac{\beta_1^2\gamma_{m0}^2}{\beta_0^4}+ \frac{\beta_1\gamma_{m0}\gamma_{m1}}{\beta_0^3}\right)
\right. \nonumber \\
&+& \left.\frac{\ln \xi}{\xi^2}\left(\frac{\beta_1^2\gamma_{m0}^2}{\beta_0^4}
-\frac{\beta_1\gamma_{m1}}{\beta_0^2} - \frac{\beta_1\gamma_{m0}\gamma_{m1}}{\beta_0^3}\right)+\left( \frac{\beta_1^2\gamma_{m0}}{2\beta_0^3}+\frac{\beta_1^2\gamma_{m0}^2}{2\beta_0^4}
\right)\frac{\ln^2 \xi}{\xi^2}\right]. 
\end{eqnarray}
\end{widetext}

Likewise, the solutions for the background field
at orders $\hbar^0$, $\hbar$ and $\hbar^2$, using the boundary conditions \(\bar{\varphi}_0(0) = \varphi(\mu)\) and \(\bar{\varphi}_1(0) = \bar{\varphi}_2(0) = 0\), are found to be given, respectively, by
\begin{eqnarray}\label{varphibar0}
\bar{\varphi}_0 &=&\frac{\varphi}{\xi^{-\gamma_{0}/\beta_0}},
\end{eqnarray}
\begin{eqnarray}\label{varphibar1}
\bar{\varphi}_1 &=& \frac{\lambda \varphi}{\xi^{-\gamma_{0}/\beta_0}}\left[\left(
\frac{\beta_1\gamma_{0}}{\beta_0^2}-\frac{\gamma_{1}}{\beta_0}\right)[\xi^{-1}-1]
+\frac{\beta_1\gamma_{0}}{\beta_0^2}\frac{\ln \xi}{\xi}\right], \nonumber \\
\end{eqnarray}
and
\begin{eqnarray}\label{varphibar2}
\bar{\varphi}_2 &=&\frac{\lambda^2 \varphi}{\xi^{-\gamma_{0}/\beta_0}}\left[\left( 
\frac{\beta_1^2\gamma_{0}}{2\beta_0^3}-\frac{\beta_2\gamma_{0}}{2\beta_0^2}
+\frac{\beta_1^2\gamma_{0}^2}{2\beta_0^4}+\frac{\beta_1\gamma_{1}}{2\beta_0^2}+\frac{\gamma_{1}^2}{2\beta_0^2} \right. \right. \nonumber \\
&-& \left. \frac{\beta_1\gamma_{0}\gamma_{1}}{\beta_0^3}
-\frac{\gamma_{2}}{2\beta_0}\right)[\xi^{-2}-1] + \left(-\frac{\beta_1^2\gamma_{0}}{\beta_0^3}+\frac{\beta_2\gamma_{0}}{\beta_0^2}\right. \nonumber\\
&-&\left. \frac{\beta_1^2\gamma_{0}^2}{\beta_0^4}
+\frac{2\beta_1\gamma_{0}\gamma_{1}}{\beta_0^3}-\frac{\gamma_{1}^2}{\beta_0^2}\right)[\xi^{-1}-1] + \left(-\frac{\beta_1^2\gamma_{0}^2}{\beta_0^4} \right. \nonumber\\
&+& \left. \frac{\beta_1\gamma_{0}\gamma_{1}}{\beta_0^3}\right)
\frac{\ln \xi}{\xi}+\left(\frac{\beta_1^2\gamma_{0}^2}{\beta_0^4}
+\frac{\beta_1\gamma_{1}}{\beta_0^2}-\frac{\beta_1\gamma_{0}\gamma_{1}}{\beta_0^3}\right)
\frac{\ln \xi}{\xi^2}\nonumber\\
&+&\left. \left( -\frac{\beta_1^2\gamma_{0}}{2\beta_0^3}+\frac{\beta_1^2\gamma_{0}^2}{2\beta_0^4}
\right)\frac{\ln^2 \xi}{\xi^2}\right].
\end{eqnarray}

{}Finally, the solutions for the vacuum energy at orders $\hbar^0$, $\hbar$ and $\hbar^2$, using the boundary conditions \(\bar{\Lambda}_0(0) = \bar{\Lambda}_1(0) = \bar{\Lambda}_2(0) = 0\),  are found to be given, respectively, by
\begin{eqnarray}\label{Lambdabar0} 
    \bar{\Lambda}_0 &=& -\frac{m^4\beta_{\Lambda 0}}{\lambda(\beta_0-2\gamma_{m0})}[\xi^{1-2\gamma_{m0}/\beta_0}-1],
\end{eqnarray}
\begin{eqnarray}\label{Lambdabar1} 
\bar{\Lambda}_1 &=& m
^4\left[\frac{2\beta_{\Lambda 0}}{\beta_0(\beta_0-2\gamma_{m0})}\left(
-\frac{\beta_1\gamma_{m0}}{\beta_0}+\gamma_{m1}\right) \right. \nonumber \\
&\times& [\xi^{1-2\gamma_{m0}/\beta_0}-1] -\left(\frac{\beta_1\beta_{\Lambda 0}}{\beta_0^2}+\frac{\beta_1\beta_{\Lambda 0}}{2\beta_0\gamma_{m0}} \right. \nonumber \\
&-& \left. \frac{\beta_{\Lambda 0}\gamma_{m1}}{\beta_0\gamma_{m0}}-\frac{\beta_{\Lambda 1}}{2\gamma_{m0}}\right)
[\xi^{-2\gamma_{m0}/\beta_0}-1]\nonumber\\
&-&\left. \frac{\beta_{\Lambda 0}\beta_1}{\beta_0^2}\frac{\ln \xi}{\xi^{2\gamma_{m0}/\beta_0}}\right]\;,
\end{eqnarray}
and
\begin{eqnarray}\label{Lambdabar2} 
\bar{\Lambda}_2 &=& \lambda m^4 \left[\left(
\frac{\beta_1\beta_{\Lambda 1}}{\beta_0^2}
-\frac{\beta_2\beta_{\Lambda 0}}{\beta_0^2}-\frac{2\beta_1^2\beta_{\Lambda 0}\gamma_{m0}}{\beta_0^4} \right. \right. \nonumber\\
&+&\frac{4\beta_1\beta_{\Lambda 0}\gamma_{m1}}{\beta_0^3}-\frac{\beta_{\Lambda 1}\gamma_{m1}}{\beta_0\gamma_{m0}}+\frac{\beta_1\beta_{\Lambda 0}\gamma_{m1}}{\beta_0^2\gamma_{m0}} \nonumber \\
&-& \left. \frac{2\beta_{\Lambda 0}\gamma_{m1}^2}{\beta_0^2\gamma_{m0}}\right)
[\xi^{-2\gamma_{m0}/\beta_0}-1] + \left(\frac{\beta_1^2\gamma_{m0}}{\beta_0^2} \right.\nonumber\\
&-&\frac{\beta_2\gamma_{m0}}{\beta_0}-\frac{2\beta_1^2\gamma_{m0}^2}{\beta_0^3}+{4\beta_1\gamma_{m0}\gamma_{m1}}{\beta_0^2}+\gamma_{m2}\nonumber\\
&-&\left. {2\gamma_{m1}^2}{\beta_0} -\frac{\beta_1\gamma_{m1}}{\beta_0} \right)\frac{\beta_{\Lambda 0} (\xi^{1-2\gamma_{m0}/\beta_0}-1)}{\beta_0(\beta_0-2\gamma_{m0})}\nonumber\\
&+&\left(\beta_{\Lambda 2}-\frac{\beta_1\beta_{\Lambda 1}}{\beta_0}-\frac{2\beta_1\beta_{\Lambda 1}\gamma_{m0}}{\beta_0^2}+\frac{\beta_1^2\beta_{\Lambda 0}\gamma_{m0}}{\beta_0^3} \right. \nonumber\\
&+&\frac{\beta_2\beta_{\Lambda 0}\gamma_{m0}}{\beta_0^2}+\frac{2\beta_1^2\beta_{\Lambda 0}\gamma_{m0}^2}{\beta_0^4}+\frac{2\beta_{\Lambda 1}\gamma_{m1}}{\beta_0}\nonumber\\
&-& \frac{3\beta_1\beta_{\Lambda 0}\gamma_{m1}}{\beta_0^2}-\frac{4\beta_1\beta_{\Lambda 0}\gamma_{m0}\gamma_{m1}}{\beta_0^3}+\frac{2\beta_{\Lambda 0}\gamma_{m1}^2}{\beta_0^2} \nonumber \\
&+& \left.\frac{\beta_{\Lambda 0}\gamma_{m2}}{\beta_0}\right)\frac{(\xi^{-1-2\gamma_{m0}/\beta_0}-1)}{\beta_0+2\gamma_{m0}}\nonumber\\
&+&\left(\frac{2\beta_1\beta_{\Lambda 0}\gamma_{m1}}{\beta_0^3}-\frac{2\beta_1^2\beta_{\Lambda 0}\gamma_{m0}}{\beta_0^4}\right)
\frac{\ln \xi}{\xi^{2\gamma_{m0}/\beta_0}}\nonumber\\
&+&\frac{\ln \xi}{\xi^{1+2\gamma_{m0}/\beta_0}}\left(-\frac{\beta_1\beta_{\Lambda 1}}{\beta_0^2}+\frac{2\beta_1^2\beta_{\Lambda 0}\gamma_{m0}}{\beta_0^4} \right. \nonumber \\
&-& \left.\frac{2\beta_1\beta_{\Lambda 0}\gamma_{m1}}{\beta_0^3}\right)
+\left. \frac{\beta_1^2\beta_{\Lambda 0}\gamma_{m0}}{\beta_0^4}\frac{\ln^2 \xi}{\xi^{1+2\gamma_{m0}/\beta_0}} \right].
\end{eqnarray}

The solutions for the coupling, mass, background field, and vacuum energy at $\hbar^3$ can be found in the Ref.~\cite{Chung:1999xm}.

\section{VRG Effective Potential}\label{app:vrgdelta2}

Here we give the complete expressions for the VRG effective potential at orders $\delta$ and $\delta^2$ that were used in our numerical studies.

The VRG effective potential at order $\delta$ is given by
\begin{eqnarray}\label{VRG1}
    V_{\rm VRG}^{\delta} &=& \frac{1}{2} \bar{\Omega}^2\varphi^2+\frac{1}{4!} \frac{\lambda}{\xi} \varphi^4 - \frac{\bar{\Omega}^{4}\left( 2 L_{\bar{\mu}} +3\right)}{8\left( 4\pi \right) ^{2}} \nonumber \\
    &-&\frac{J_{0,\bar{\Omega}} T^{4}}{2\left( 4\pi \right)^{2} } -\frac{\lambda }{8\left( 4\pi \right) ^{4} \xi}\left[ \left( L_{\bar{\mu}} +1\right)^2 \bar{\Omega}^{4}-J_{1,\bar{\Omega}}^2 T^{4}\right] \nonumber \\
    &+& \frac{ \lambda \varphi^2}{4(4\pi )^{2} \xi} J_{1,\bar{\Omega}} T^{2} + \frac{\eta ^{2}}{2\left( 4\pi \right) ^{2}}\left[ \left(L_{\bar{\mu}} +1\right) \bar{\Omega}^{2} \right. \nonumber \\
    &-& \left. J_{1,\bar{\Omega}} T^{2}\right] + F_{\rm vac , 1},
\end{eqnarray}
where $\xi = 1 - \lambda \beta_0 t$, $t = \ln(\bar{\mu}/\mu)$ and we have also defined
\begin{eqnarray}
\bar{\Omega}^2 &=& \frac{1}{\xi^{1/3}}(m^2+ \eta^2), 
 \\
L_{\bar{\mu}} &=& \ln(\mu^2 e^{2t}/\bar{\Omega}^2),
\end{eqnarray}
and
\begin{widetext}
\begin{eqnarray}
    F_{\rm vac , 1} &=& \frac{ \left(\eta ^2-\xi^{1/3} \bar{\Omega} ^2\right)^2}{2 \lambda }\left(1-\xi^{1/3}\right)
    +\frac{\bar{\Omega}^4}{54(4\pi)^2} \left[35 - 54 \xi^{2/3} + 17 \ln \xi+19 \xi\right] \nonumber \\
    &+& \frac{\bar{\Omega}^2 \lambda}{108 \xi (4\pi)^4}\left[19 (\xi -1)-34 \ln \xi \right]\left[ \left( L_{\bar{\mu}} +1\right) \bar{\Omega}^{2}-J_{1,\bar{\Omega}} T^{2}\right] 
    \nonumber \\
    &-&\frac{\bar{\Omega}^2 \lambda \varphi ^2 
    \left[ 8 (\xi-1) -17 \ln \xi\right]}{54 \xi  (4\pi)^2}- \frac{\bar{\Omega}^4 \lambda}{\xi (4\pi)^4} \left[-\frac{289 \ln ^2\xi}{1458}+   \frac{7776 \zeta (3)+2509}{11664} \xi ^2                  \right.
    \nonumber \\
    &+& \left. 
    \frac{323 (\xi -1) \ln \xi }{1458} -\frac{36 \zeta (3)+23}{30} \xi ^{5/3}- \frac{3888 \zeta (3)+9239}{29160} +\frac{7776 \zeta (3)+10129}{11664}\xi \right].
    \nonumber \\
\end{eqnarray}

At order $\delta^2$, the VRG effective potential is given by
\begin{eqnarray}\label{VRG2}
    V_{\rm VRG}^{\delta^2} &=& V_{\rm VRG}^{\delta} + \frac{\eta ^4}{4\xi^{2/3} (4\pi)^2} \left(L_{\bar{\mu}}+J_{2,\bar{\Omega} }\right) + \frac{\lambda \eta^2}{4\xi^{4/3}\left( 4\pi \right) ^{4}}\left[ J_{1,\bar{\Omega}}J_{2,\bar{\Omega}} T^{2} + L_{\bar{\mu}}\left( L_{\bar{\mu}}+1\right) \bar{\Omega}^{2}\right] 
    \nonumber \\
    &-& \frac{\lambda^2}{48 \xi^2 \left( 4\pi \right) ^{6}}\left[\left( 5 L_{\bar{\mu}}^{3} 
    + 7L_{\bar{\mu}}^{2} +\left(\frac{57}{2} + 4C_1 + \pi^2\right)L_{\bar{\mu}}-  5-\frac{11 \pi^2}{12} - \psi ^{\prime \prime }\left( 1\right) + C_0 + 4C_{1} \right) \bar{\Omega}^{4} 
     \right. \nonumber \\
    &+& \left. \left( 2 J_{1,\bar{\Omega}}^{2} +J_{1,\bar{\Omega}}^{2} J_{2,\bar{\Omega}}+ 2K_{2,\bar{\Omega}}+\frac{4}{3}K_{3,\bar{\Omega}}\right) 3 T^{4} \right] 
    \nonumber \\
    &+& \varphi^2 \left\{ \frac{\lambda \eta ^{2}}{4\xi^{4/3}(4\pi) ^{2}}J_{2,\bar{\Omega}} 
    + \frac{ \lambda ^{2}}{4\xi^2(4\pi) ^{4}} \left[-T^2 \left( \frac{1}{2}  J_{1,\bar{\Omega}}  J_{2,\bar{\Omega}}+ H_{2,\bar{\Omega}} + H_{3,\bar{\Omega}} 
    -   J_{1,\bar{\Omega}}\right) 
    \right.\right.
    \nonumber \\
    &+& \left. \left. \left(\frac{C_1}{3}+C_2+\frac{L_{\bar{\mu}}}{6}\right)\bar{\Omega} ^2 \right] \right\}
   + \frac{\lambda ^{2} \varphi^4}{48 \xi^2 (4\pi) ^{2}} (2-3 J_{2,\bar{\Omega}})+  F_{\rm vac, 2} + F_{\varphi},
\end{eqnarray}
where
%
\begin{eqnarray}
    F_\varphi &=& \varphi^2 \left\{ \frac{\lambda^2 \bar{\Omega}^2 (19 (\xi -1)-34 \ln \xi) }{216 \xi ^{2} (4\pi)^4}\left(L_{\bar{\mu}} + J_{2,\bar{\Omega}}\right)  - \frac{\lambda^2(\xi +34 \ln \xi -1)}{72 \xi ^2 (4\pi)^4}\left[(L_{\bar{\mu}}+1) \bar{\Omega} ^2-T^2 J_{1,\bar{\Omega}}\right] \right. \nonumber \\
    &+& \left. \frac{\lambda^2 \bar{\Omega}^2}{\xi ^{2} (4\pi)^4} \left[\frac{17 (43 + 108L_{\bar{\mu}} - 16 \xi) \ln \xi }{5832} + \frac{289 \ln ^2\xi }{729} 
    + \frac{(\xi -1)}{11664}(-864L_{\bar{\mu}} +973 \xi + 3888\zeta(3) (\xi-1) - 4186) \right] \right. \nonumber \\
    &+& \left. \frac{\eta ^2 \lambda  (8 (\xi -1) -17 \ln \xi )}{54 \xi ^{4/3} (4\pi)^2} \right\} +  \frac{\lambda^2 \varphi^4\left(\xi -1 +17 \ln\xi\right)}{216 \xi ^2 (4\pi)^2},
    \nonumber \\
\end{eqnarray}
and
%
%
\begin{eqnarray}
    F_{\rm vac,2} &=& \frac{\eta^4\left(35 - 54 \xi^{2/3} +19 \xi + 17 \ln\xi \right)}{54(4\pi)^2 \, \xi^{2/3}} +\frac{17 \lambda ^2 \ln \xi}{72 \xi ^2 (4\pi)^6} \left[(L_{\bar{\mu}}+1) \bar{\Omega} ^2-T^2 J_{1,\bar{\Omega}}\right]^2 + \frac{\eta ^2 \lambda  \bar{\Omega} ^2}{\xi ^{4/3} (4\pi)^4} \left[-\frac{23 \xi ^{5/3}}{15}+\frac{2509 \xi ^2}{5832} \right. \nonumber \\
    &+& \left. \frac{10129 \xi }{5832}-\frac{289 \ln ^2\xi }{729}+\frac{323}{729} \xi  \ln \xi -\frac{323 \ln \xi }{729}-\frac{1}{5} 12 \xi ^{5/3} \zeta (3)+\frac{4 \xi ^2 \zeta (3)}{3}+\frac{4 \xi  \zeta (3)}{3}-\frac{4 \zeta (3)}{15}-\frac{9239}{14580}\right]\nonumber \\
    &+& \frac{\lambda ^2 \bar{\Omega}^2 (19 (\xi -1)-34 \ln \xi)^2}{23328 \xi ^2 (4\pi)^6}\left\{\left[(L_{\bar{\mu}}+1) \Omega ^2-T^2 J_{1,\bar{\Omega}}\right]-2 \bar{\Omega} ^2 \left(L_{\bar{\mu}}+J_{2,\bar{\Omega}}\right) \right\} +\left\{\left[(L_{\bar{\mu}}+1) \Omega ^2-T^2 J_{1,\bar{\Omega}}\right] \right. \nonumber \\
    &-& \left. 2 \bar{\Omega} ^2 \left(L_{\bar{\mu}}+J_{2,\bar{\Omega}}\right) \right\} \frac{\lambda ^2 (L_{\bar{\mu}}+1) \bar{\Omega} ^2 (19 (\xi -1)-34 \ln \xi )}{432 \xi ^2 (4\pi)^6} -\frac{\lambda ^2 \bar{\Omega} ^2 (19 (\xi -1)-34 \ln (\xi ))}{216 \xi ^2 (4\pi)^6}\left(L_{\bar{\mu}}+J_{2,\bar{\Omega}}\right) \nonumber \\
    &\times & \left[(L_{\bar{\mu}}+1) \bar{\Omega} ^2-T^2 J_{1,\bar{\Omega}}\right] +\frac{\lambda ^2 \bar{\Omega} ^2}{18895680 \xi ^2 (4\pi)^6} \left\{-4860 T^2 J_{1,\bar{\Omega}} \left[-34 (63 L_{\bar{\mu}}-19 \xi +34) \ln \xi -1734 \ln ^2\xi  \right. \right. \nonumber \\
    &+& \left. (\xi -1) (171 L_{\bar{\mu}}-2 \xi  (648 \zeta (3)+179)+1296 \zeta (3)+1435) \right] + \bar{\Omega}^2 \left[831060  L_{\bar{\mu}}^2 (\xi -1)-981264 \xi +5923585 \right. \nonumber \\
    &-& 34680 (243 L_{\bar{\mu}}+76 \xi +257) \ln ^2\xi -314928 \pi ^4 \xi ^{8/3}+83154114 \xi ^{8/3}+209952 \pi ^4 \xi ^3-33518530 \xi ^3\nonumber \\
    &+& 104976 \pi ^4 \xi ^2-54577905 \xi ^2+3930400 \ln ^3\xi -9720 L_{\bar{\mu}} (\xi -1) (\xi  (648 \zeta (3)+179)-648 \zeta (3)-803)\nonumber \\
    &+& 108177768 \xi ^{8/3} \zeta (3)+141717600 \xi ^{8/3} \zeta (5)-67301280 \xi ^3 \zeta (3)-83980800 \xi ^3 \zeta (5)-45839520 \xi ^2 \zeta (3)\nonumber \\
    &-&62985600 \xi ^2 \zeta (5)+1959552 \xi  \zeta (3)+3003480 \zeta (3)+5248800 \zeta (5)-1020 \ln \xi  \left(10206 L_{\bar{\mu}}^2-162 L_{\bar{\mu}} (19 \xi -97) \right.\nonumber \\
    &-& \left.\left.\left.\xi ^2 (7776 \zeta (3)+2509)+4 \xi  (3888 \zeta (3)+2357)-7776 \zeta (3)-5218\right)\right] \right\}.
\end{eqnarray}
\end{widetext}


\end{document}